# Atomically Modulating Competing Exchange Interactions in Centrosymmetric Skyrmion Hosts $GdRu_2X_2$ (X = Si, Ge)

Dasuni N. Rathnaweera,[1] Xudong Huai,[1] K. Ramesh Kumar,[1] Michał J. Winiarski,[2] Tomasz Klimczuk,[2] Allana G. Iwanicki,[3,4] Satya Kushwaha,[3,4] Martin Mourigal[5], Tyrel M. McQueen,[3,4,6] and Thao T. Tran[1,*]

[1]Department of Chemistry, Clemson University, Clemson, South Carolina, 29634, USA

[2]Faculty of Applied Physics and Mathematics and Advanced Materials Center, Gdansk University of Technology, ul. Narutowicza 11/12, 80-233 Gdansk, Poland

[3]Institute for Quantum Matter, William H. Miller III Department of Physics and Astronomy, Johns Hopkins University, 3400 N. Charles Street, Baltimore, MD 21218, USA

[4]Department of Chemistry, Johns Hopkins University, 3400 N. Charles Street, Baltimore, MD, 21218, USA

[5]School of Physics, Georgia Institute of Technology, Atlanta, GA, 30332, USA

[6]Department of Materials Science and Engineering, Johns Hopkins University, Baltimore, 21218, MD, USA

**ABSTRACT:** Magnetic skyrmions–topologically protected spin states–can enable high-density, low-power spintronics. Despite the growth of interest in skyrmion research, the microscopic mechanisms leading to skyrmion phase transitions at specific temperatures and magnetic fields remain elusive. In this work, we systematically study the isostructural centrosymmetric magnets—$GdRu_2X_2$ (X = Si and Ge)—and the role of X-$p$ orbitals in modifying magnetic exchange interactions. Electronic structure and exchange interaction evaluations reveal that the more extended Ge-4$p$ vs. Si-3$p$ orbitals enhance competing exchange interactions in $GdRu_2Ge_2$, thereby manifesting skyrmions evolution condition in $GdRu_2X_2$. $GdRu_2Ge_2$ single crystals exhibit two high-entropy regions associated with skyrmion phases at 0.9 T ≤ $\mu_0H$ ≤ 1.2 T and 1.3 T ≤ $\mu_0H$ ≤ 1.7 T, 2 K ≤ $T$ ≤ 30 K—lower field and higher temperature than that in the Si counterpart. The positive entropy change associated with the skyrmion phase transition is extracted from heat capacity and magnetization data. Transport measurements reveal a topological Hall effect, validating the nontrivial spin textures and Berry curvature. This work demonstrates, for the first time, how modifying exchange interactions at the atomic level enables the tunability of skyrmion phase transitions, enabling a significant step toward developing skyrmions at desired temperatures and fields.

## INTRODUCTION

Understanding the chemistry and physics underlying the evolution of topologically distinct spin textures—magnetic skyrmions—and their behaviors is essential for advancing future information and energy technologies.[1] Magnetic skyrmions, displaying a unique, dynamic quasiparticle character with a size ranging from a few to ~ 100 nm, can enable high-density, low-power spintronics and logic functions.[2] The topological protection ensures skyrmions retain their unique spin properties, even in the presence of defects and disorders, which are inevitable in real materials. The formation of magnetic skyrmions is mainly driven by two fundamental mechanisms based on crystal symmetries. In noncentrosymmetric magnets, the antisymmetric Dzyaloshinskii-Moriya interaction, facilitated by large spin−orbit coupling, can stabilize skyrmion formation.[3] In centrosymmetric magnetic metals, the long-range Ruderman-Kittel-Kasuya-Yosida (RKKY) exchange interaction $J(r) \sim \sin(2k_F r)/r^3$, where $k_F$ is the Fermi wavevector of conduction electrons and $r$ is the distance between the magnetic moments, assisted by magnetic frustration, facilitates skyrmion evolution.[2c, 4] Enhanced competing ferromagnetic (FM) and antiferromagnetic (AFM) interactions at a similar energy scale promote the emergence of skyrmions.

Small-sized skyrmions in centrosymmetric magnetic metals often display higher topological charge density compared to those in noncentrosymmetric counterparts. This feature is a key for enhancing the topological Hall effect—a physical phenomenon with significant implications for advanced spintronic technologies.[4c],[2c, 5] Centrosymmetric skyrmion hosts are limited to a few metals adopting cubic, hexagonal, and tetragonal crystal lattices with $Gd^{3+}$ and $Eu^{2+}$ ($S$ = 7/2, $L$ = 0) spins, such as $Gd_2PdSi_3$, $Gd_3Ru_4Al_{12}$, $EuAl_4$, and $EuGa_2Al_2$.[4d, 6] Recent theoretical studies on centrosymmetric tetragonal lattices suggested several factors influencing skyrmion formation, such as frustrated exchange interactions, easy-axis anisotropy, bond-dependent anisotropy, and positive biquadratic interactions.[7] While these studies presented tremendous progress in skyrmion materials research, a picture of how competing exchange interactions can be modified at the atomic and molecular levels remains unclear. This is a daunting task due to the intricate nature of skyrmion hosts and phase transitions to skyrmions at a given field and temperature.

$GdRu_2Si_2$ was found to display a skyrmion region at 2 T ≤ $\mu_0 H$ ≤ 2.5 T, 2 K ≤ $T$ ≤ 20 K, with the smallest diameter of 1.9 nm among the known skyrmion materials.[8] It has been demonstrated that indirect RKKY interactions stabilize equivalent magnetic modulation vectors, resulting in a square skyrmion lattice in the Si material.[8a, 9] $GdRu_2Ge_2$—an isostructural sibling—has recently been realized to feature the successive formation of two skyrmion phases at 0.9 T ≤ $\mu_0 H$ ≤ 1.2 T and 1.3 T ≤ $\mu_0 H$ ≤ 1.7 T, 2 K ≤ $T$ ≤ 30 K.[10] Studies showed that this rich topological behavior is governed by the presence of competing RKKY exchange interactions at inequivalent wavevectors.[10] Realizing how to achieve zero-field skyrmion formation and how magnon dispersions can be linked to lower-field magnetism are critical long-term goals for skyrmion research and their diverse applications. While $GdRu_2X_2$ (X = Si and Ge) presents an exciting system for tuning skyrmion phase transitions, *how* the exchange interactions play a role in stabilizing skyrmions in $GdRu_2X_2$ and the underlying reasons that lead to different critical field and temperature conditions at which the skyrmion phase transitions take place are not yet understood.

To address multifaceted research objective, our work focused on understanding the role of the X-$p$ orbitals in modifying the Gd—Gd magnetic exchange interactions in $GdRu_2X_2$ and whether and/or how this interplay contributes to the critical temperature and field of the phase transition to skyrmions. We first conducted an analysis of magnetic exchange interactions in conjunction with spin-polarized band structure and density of states with functional theory (DFT) calculations. We then supplemented our results with experimental evidence. We synthesized $GdRu_2Ge_2$ crystals and characterized their structural, magnetic, thermodynamic, and transport properties. The integrated approach enables us to gain insights into the impact of Si-3$p$/Ge-4$p$ orbitals on the Gd—Gd magnetic exchange interactions while linking this interplay to the skyrmion phase transitions in $GdRu_2X_2$.

**RESULTS AND DISCUSSION**

$GdRu_2X_2$ (X = Si and Ge) adopts the $ThCr_2Si_2$-type structure—a centrosymmetric tetragonal space group $I4/mmm$. $ThCr_2Si_2$-type is the most ubiquitous structural arrangement for ternary element combinations, allowing for pronounced chemical flexibility.[11] Particularly, when $f$-elements are incorporated into this chemical framework, a diverse spectrum of emergent properties is observed, including complex magnetism, charge instabilities, structural instabilities, unconventional superconductivity, and topologically protected electronic states.[9, 12] Both $GdRu_2Si_2$ and $GdRu_2Ge_2$ host skyrmions, but at different critical field and temperature conditions. To understand the underlying physical origins that give rise to this contrast, we calculated magnetic exchange interactions of $GdRu_2X_2$ using the Quantum Espresso Software package[13]. We performed an energy-mapping analysis as illustrated in **Figure 1**.[14] The total energies of the different possible spin configurations were calculated, and these energies were then "mapped" onto the spin Hamiltonian, allowing the spin exchange parameters to be extracted quantitatively. The spin Hamiltonian $\hat{H}_{\text{spin}}$ is defined in terms of several different spin exchange parameters as follows:[14b]

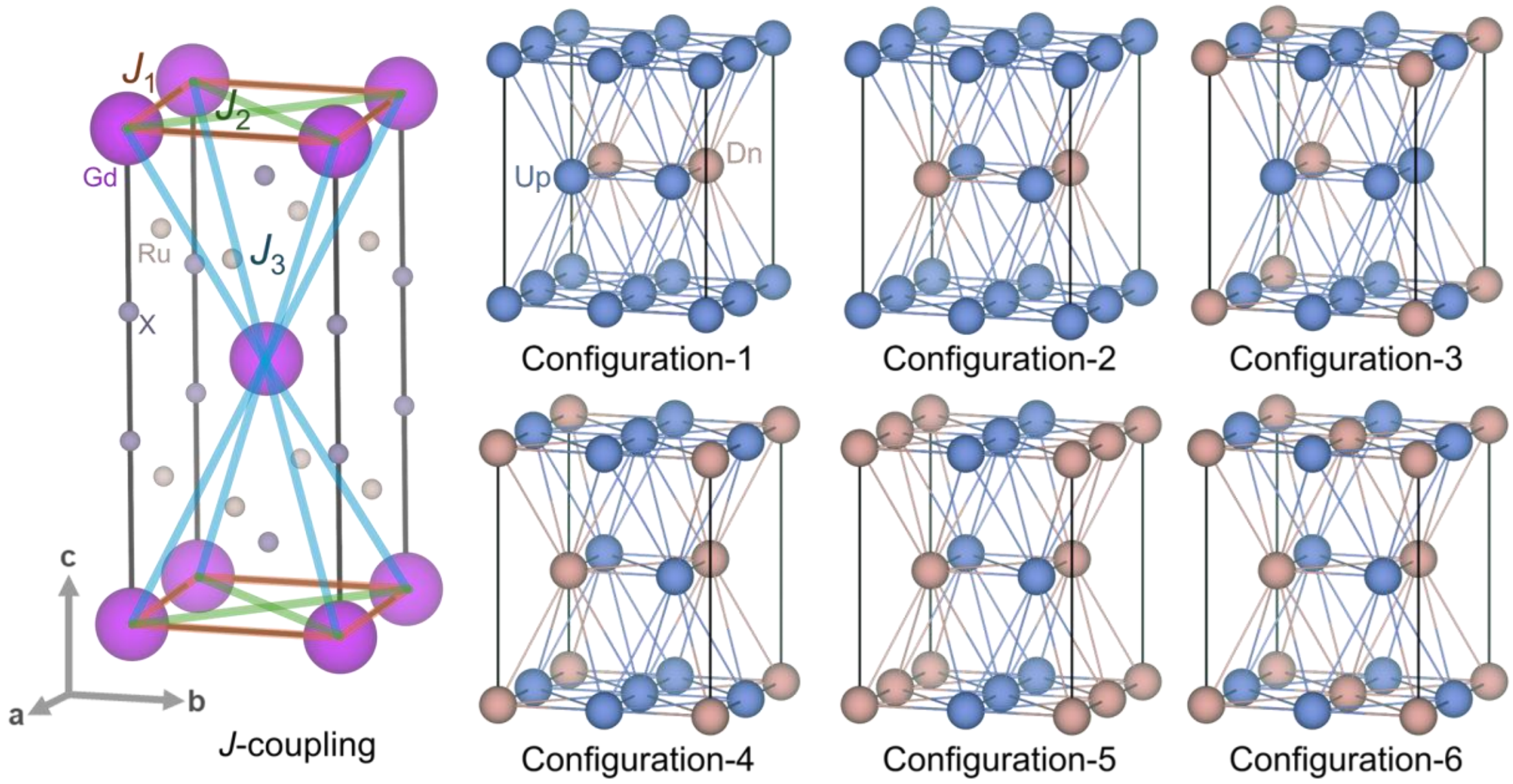

**Figure 1.** Representing exchange interactions $J_1$, $J_2$ and $J_3$ between $Gd^{3+}$ atoms within the unit cell of $GdRu_2X_2$ (X =Si and Ge) and spin-ordered states within the (2a, 2b, c) supercell. Pink and blue colors correspond to the spin up and down, respectively.

$$\hat{H}_{spin} = -\sum_{<ij>} J_{ij} \hat{S}_i \, . \, \hat{S}_j \qquad (1)$$

where $\hat{S}_i$ and $\hat{S}_j$ are the spins at the sites *i* and *j,* and $J_{ij}$ is the exchange parameter describing the sign and strength of the interaction between the spin sites *i* and *j.* Equation (1) sums up all magnetic interactions within $GdRu_2X_2$. The Gd spins form two square sublattices within a unit cell. Our model incorporates the nearest-neighbor exchange interaction $J_1$ along the *a*- or *b*-axis, the next nearest-neighbor exchange interaction $J_2$ within the *ab*-plane along the [110] direction, and the interaction $J_3$ between the Gd square sublattices. To accurately simulate an extended solid and avoid artificial interactions between periodic images of atoms, we used a (2*a*, 2*b*, *c*) supercell (containing four formula units) and six different spin-ordered states (**Figure 1**). The spin-polarized DFT calculations were performed using the Quantum Espresso software package.[14d] The total spin exchange energy per supercell can be expressed by the following equations:[15]

$$\begin{aligned} E_1 &= E_0 + (-8J_1 + 0J_2 + 0J_3) \cdot S^2 \qquad (2) \\ E_2 &= E_0 + (0J_1 - 16J_2 + 0J_3) \cdot S^2 \\ E_3 &= E_0 + (0J_1 + 0J_2 - 8J_3) \cdot S^2 \\ E_4 &= E_0 + (8J_1 - 8J_2 + 0J_3) \cdot S^2 \\ E_5 &= E_0 + (8J_1 + 0J_2 + 0J_3) \cdot S^2 \end{aligned}$$

$$E_6 = E_0 + (16J_1 - 16J_2 + 0J_3) \cdot S^2$$

where the $E_0$ corresponds to the non-magnetic contribution to the total energy, and $S$ = 7/2, the spin for $Gd^{3+}$. From these energies, the exchange interactions per four formula units can be calculated as:

$$J_1 = \frac{(E_6 - E_2)}{16S^2} \quad (3)$$
$$J_2 = \frac{(E_5 - E_4)}{8S^2}$$
$$J_3 = \frac{(E_1 + E_5 - 2E_3)}{16S^2}$$

The $J$-coupling interaction can then be obtained:

**Table 1.** Calculated $J$-coupling constants for $GdRu_2X_2$ (X = Si, and Ge)

| X Site | $J_1$ (meV) | $J_2$ (meV) | $J_3$ (meV) |
|---|---|---|---|
| **Si** | -0.06(2) | -13.22(2) | -0.04(2) |
| **Ge** | 6.12(2) | -13.24(2) | -0.20(2) |

As shown in **Table 1**, $J_1$ varies from negligible AFM to appreciable FM when the X site changes from Si to Ge. Meanwhile, strong AFM interaction $J_2$ remains effectively unaffected by the X site. $J_3$ reveals negligible AFM for $GdRu_2Si_2$ and weak AFM for $GdRu_2Ge_2$. Overall, strong AFM $J_2$ dominates the exchange interactions in the Si compound, while those in the Ge material are more complex. The strong FM $J_1$ in $GdRu_2Ge_2$ leads to more competing AFM-FM exchange interactions in the Ge material compared to the Si sibling. This observation is intriguing, as substituting Si with Ge—an adjacent neighbor in the same group 14 in the periodic table—yields a profound modification in competing magnetic exchange interactions at similar energy scales.

To understand how this observation aligns with the physical properties of the Si and Ge compounds and how modifying exchange interactions affect the temperature and magnetic field conditions at which skrymion phase transition occurs, we investigated magnetic, thermodynamic and transport properties of $GdRu_2Ge_2$.

$GdRu_2Ge_2$ polycrystalline material was synthesized using the arc melting technique and used that polycrystalline materials to grow sizable crystals using tilted LASER diode floating zone furnace (experimental section). Our single-crystal X-ray diffraction measurements confirmed that $GdRu_2Ge_2$ adopts the centrosymmetric tetragonal space group $I4$/mmm (**Figure 2, Table S1-2**). The structure of $GdRu_2Ge_2$ consists of Gd square nets connected to [$Ru_2Ge_2$] layers. The Gd—Gd, Ru—Ge, and Gd—Ge bond distances are 4.23, 2.42, and 3.26 Å, respectively (**Table S3**). Our powder Rietveld refinement agrees well with the structure determined from the single crystal data (**Figure 2b**). Scanning Electron Microscopy (SEM) coupled with Energy Dispersive Spectroscopy (EDS) (**Figure S1**) analysis proved the chemical composition of $GdRu_2Ge_2$. Floating zone grown crystals were aligned using Laue diffractometer (**Figure 2c**). We observe the formation of two skyrmion phases in $GdRu_2Ge_2$ in the magnetoentropic mapping at 2 K ≤ $T$ ≤ 30 K, 0.9 T ≤ $\mu_0 H$ ≤ 1.2 T and 1.3 T ≤ $\mu_0 H$ ≤ 1.7 T (**Figure 3**), consistent with the work reported by Yoshimochi et al.[10] For neutron scattering and resonant X-ray scattering of the material, the reader is invited to visit the reference.[10] Here, we provide a new understanding of *how* the Gd—Gd competing exchange interactions in $GdRu_2X_2$ influence the skyrmion phase transition conditions and supplement it with experimental thermomagnetic and transport results of $GdRu_2Ge_2$.

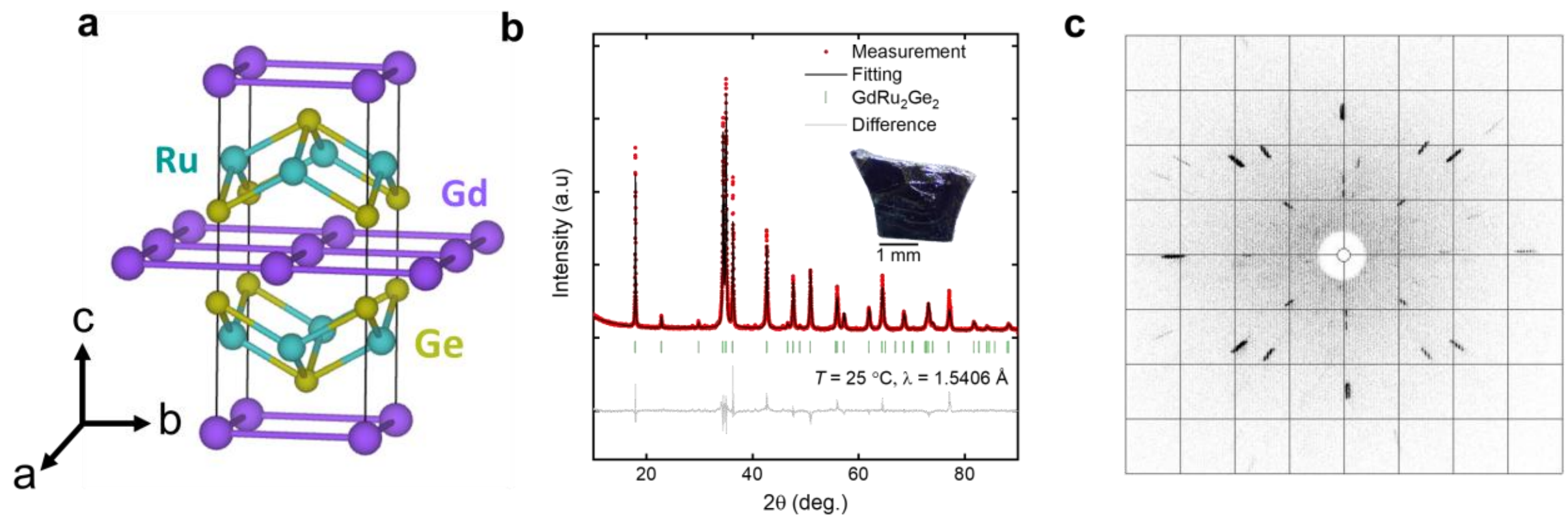

**Figure 2.** (**a**) Crystal structure of $GdRu_2Ge_2$ with the $Gd^{3+}$ square lattice, (**b**) Rietveld refinement of powder X-ray diffraction data showing phase purity of $GdRu_2Ge_2$ and inset figure showing scanning electron microscopic image of $GdRu_2Ge_2$, and (**c**) Laue diffraction pattern of [110] direction along the growth direction on a crystal from tilted-LASER diode floating zone furnace.

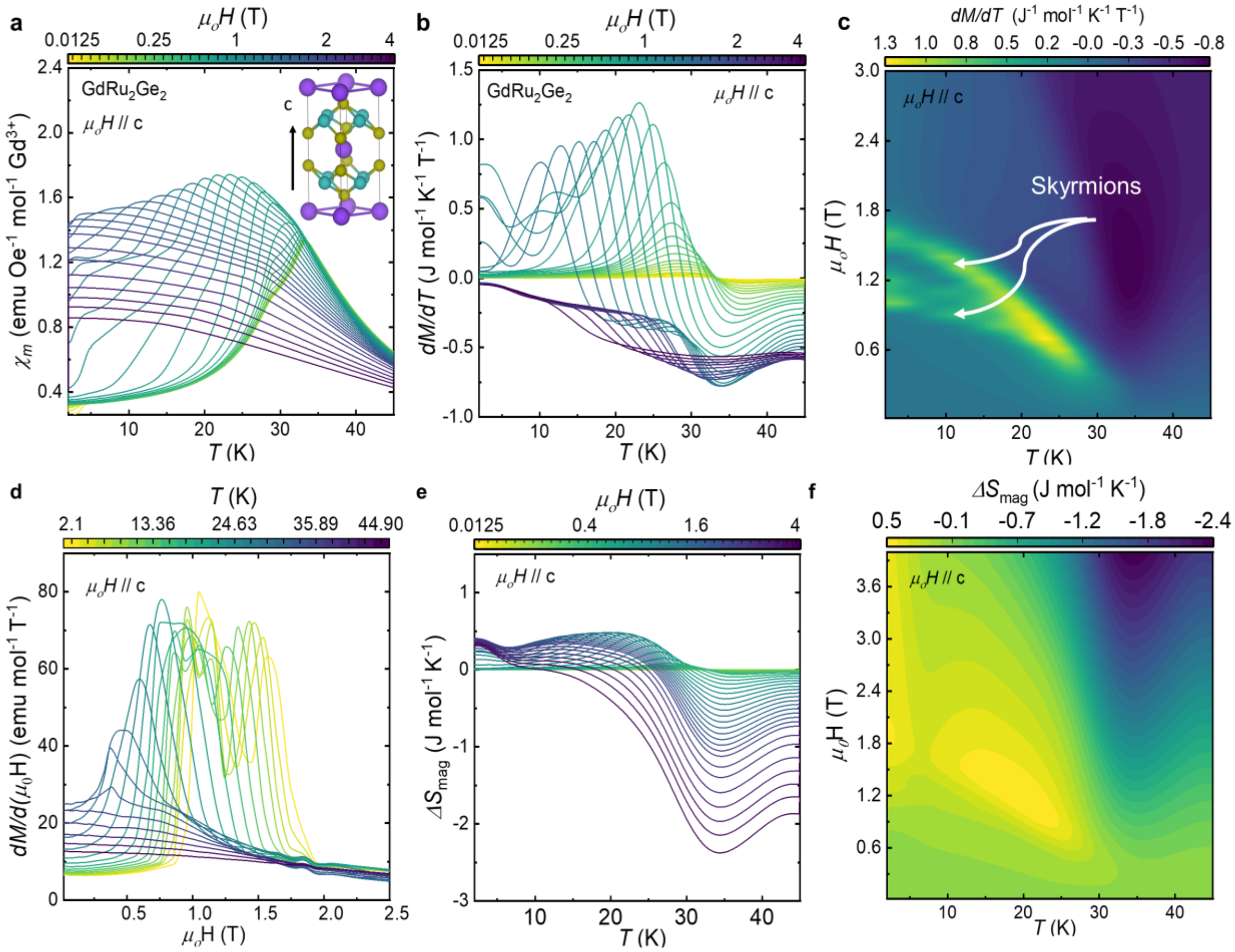

**Figure 3.** (**a**) $\chi(T)$ at different magnetic fields $\mu_0H$ // $c$, (**b**) First derivative of magnetization with respect to temperature, d$M$/d$T$ for $\mu_0H$ // $c$, (**c**) A map of d$M$/d$T$ = d$S$/d$H$ indicating first-order transitions, for $\mu_0H$//$c$, (**d**) First derivative of magnetization with respect to the magnetic field d$M$/d$H$ curve $\mu_0H$ // $c$, (**e**) Isothermal magnetic entropy at each different magnetic field obtained by the integral of d$M$/d$T$ with respect to the magnetic field $\mu_0H$ // $c$, and (**f**) A map of $\Delta S_M(T,H)$ of $GdRu_2Ge_2$.

The emergence of skyrmions typically corresponds to a positive entropy change.[16],[17],[18] The isothermal entropy change upon magnetization $\Delta S_{mag}(H,T)$ can be obtained from the Maxwell relation (Eq. 4),

$$\left(\frac{dS}{dH}\right)_T = \left(\frac{dM}{dT}\right)_H \tag{4}$$

where $S$ is the total entropy, $H$ is the magnetic field, $M$ is the magnetization, and $T$ is the temperature.

**Figure 3** depicts how the magnetization of $GdRu_2Ge_2$ evolves as a function of temperature under a series of applied fields at $\mu_0H$ // $c$ (**Figure 3a-f**) and $\mu_0H \perp c$ (**Figure S2**). The magnetic behavior of $GdRu_2Ge_2$ is orientation-dependent, confirming the magnetic anisotropy in the system as seen in previous studies.[19] A similar anisotropic behavior was observed in other skyrmion hosts with competing interactions between Gd-Gd spins.[20],[21] At $\mu_0 H \parallel c$, a rich magnetic behavior of $GdRu_2Ge_2$ is clearly observed in the first derivative of magnetization with respect to temperature $dM/dT$ curves and the magnetoentropic $\Delta S_{mag}(H, T)$ map (**Figure 3a-f**). Anomalies in the $dM/dT$ curves (**Figure 3b**) occur at the critical fields consistent with those identified in the $dM/dH$ curves (**Figure 3d**), confirming the skyrmion phase transitions. The $dM/dT = dS/dH$ and $\Delta S_{mag}(H,T)$ maps of $GdRu_2Ge_2$ can be constructed from the relation (Eq. 5)

$$\Delta S_{mag}(H,T) = \int_0^H \left(\frac{dM}{dT}\right)_{H'} dH' \qquad (5)$$

In the $dM/dT$ map at $\mu_0 H \parallel c$ (**Figure 3c**), green-yellow ridges at 2 K ≤ $T$ ≤ 30 K and 0.9 T ≤ $\mu_0 H$ ≤ 1.2 T and 1.3 T ≤ $\mu_0 H$ ≤ 1.7 T denote the regions of two skyrmion phases. **Figure 3e-f** shows the $\Delta S_{mag}(H,T)$ map highlighting the regions of a positive entropy change of approximately 0.5 J $mol^{-1}$ $K^{-1}$. These high entropy regions correspond to the skyrmion formation—an entropy-driven phase transition. A positive entropy change associated with the evolution of skyrmions sets this state of matter apart from other topologically trivial states, such as helical and conical phases. At $\mu_0 H \perp c$, the magnetoentropic features associated with the skyrmion phase transitions are observed (**Figure S2**), but not as pronounced as those at $\mu_0 H \parallel c$. This divergence confirms the magnetic anisotropy of $GdRu_2Ge_2$.

The critical fields, at which the two skyrmion phases form and other magnetic phase boundaries were identified from the $M(H)$, $\chi_m(T)$ and $dM/dT$ curves (**Figure 4a-b**), are in good agreement with the results reported by Yoshimochi et al.[10] The $M(H)$ curves at $T$ =2 K reveal skyrmion and metamagnetic phase transitions before approaching the saturated moment of the FM state (~7 $\mu_B/Gd^{3+}$) at $\mu_0 H \geq 4.5$ T. This result also highlights that the skyrmion phases are present all the way from the critical temperature $T$ = 30 K down to $T$ = 2 K. It is worth noting that the emergence of skyrmions in $GdRu_2Ge_2$ occurs at higher temperature and lower field (2 K ≤ $T$ ≤ 30 K, 0.9 T ≤ $\mu_0 H$ ≤ 1.2 T and 1.3 T ≤ $\mu_0 H$ ≤ 1.7 T) than that in $GdRu_2Si_2$ (2 K ≤ $T$ ≤ 20 K, 2 T ≤ $\mu_0 H$ ≤ 2.5 T).[8a, 9]

Due to the strong in-plane correlations obtained from the abovementioned $J$-couplings calculations, the spins predominantly align within the $ab$-plane. Consequently, these in-plane spins are more responsive to magnetic fields applied along the $c$-axis, resulting in a higher magnetic susceptibility along the $c$-axis (**Figure 4d**). Our Curie-Weiss analysis and prior work in the literature consistently show a positive Curie-Weiss temperature ($\theta_{cw}$) (**Figure 4c**), suggesting the presence of FM interactions in $GdRu_2Ge_2$.[20] The competing AFM-FM behavior is illustrated by the upturns and downturns observed in the magnetic susceptibility near the transition temperature, and supported by the hysteresis in the $M(H)$ measurements.

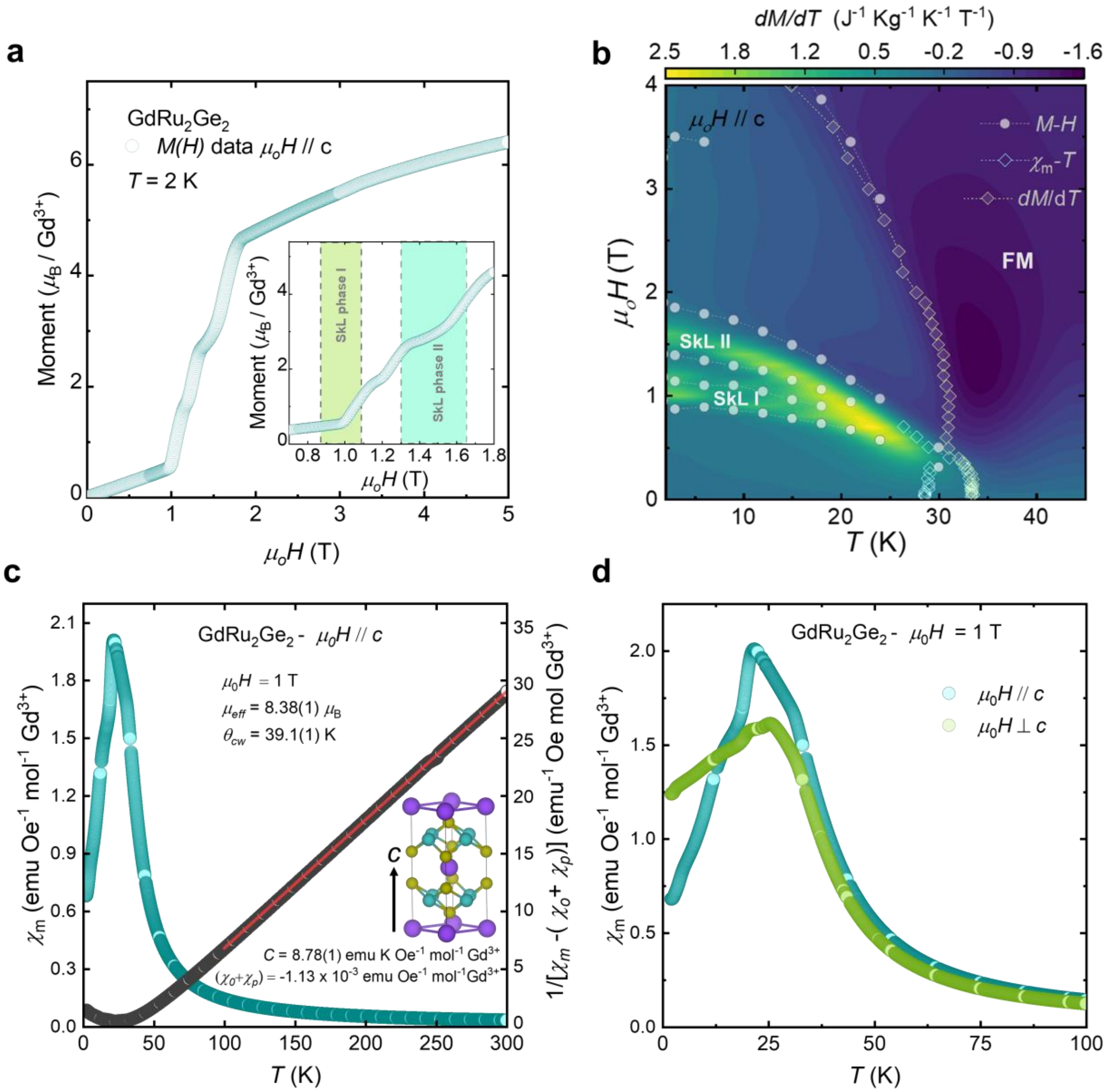

**Figure 4**. (**a**) Field dependence of the magnetization measured at $T$ = 2 K. (**b**) *B-T* magnetic phase diagram for $\mu_0 H /\!/ c$ and background contour colour map indicates variation of $dM/dT = dS/dT$, (**c**) Temperature-dependent magnetic susceptibility under constant magnetic field $\mu_0 H$ =1 T (teal) and Curie-Weiss analysis (black), and (**d**) Comparison of temperature-dependent magnetic susceptibilities at $\mu_0 H \perp c$ and $\mu_0 H /\!/ c$.

To understand the thermodynamic properties of the skyrmion host $GdRu_2Ge_2$, we performed heat capacity measurements at 2 K ≤ $T$ ≤ 300 K, $\mu_0 H$ = 0, 1.2, and 1.7 T on a crystal at $\mu_0 H /\!/ c$ (**Figure 5a**) and a polycrystalline sample (**Figure S4**). The difference between these two data sets proves the orientation dependence and magnetic anisotropy in the material. At $\mu_0 H$ = 0 T, two anomalies at $T_N$ = 32 K and $T_1$= 28 K and a hump at $T_2$ = 10 K are observed (**Figure**

**5a-b**), consistent with previous works.[19, 22] The Neel temperature of $GdRu_2Ge_2$ is also confirmed by its magnetic susceptibility data. The peak at $T_1$ could be attributed to spin re-orientation to a long-range modulated magnetic state. The hump at around $T_2 \sim 0.3T_N$ can be associated with a Schottky-like anomaly as $k_BT$ approaches the energy splitting between the two lowest-lying states due to crystal field splitting of the $^8S_{7/2}$ ground state.[23] The peak at $T_0$ signifies the transition to the skyrmion phases at $\mu_0H$ = 1.2 T, in concert with the critical fields identified from the magnetoentropic mapping and $\mathrm{d}M/\mathrm{d}H$.

To better understand field-induced metamagnetic phase transitions associated with the skyrmion formation, we analyzed the heat capacity data in more detail. In solids, lattice vibrations are typically classified into acoustic (Debye) and optical (Einstein) modes, based on the relative motion of atoms within the unit cell. Acoustic modes are in-phase vibrations where the atoms oscillate collectively, like compressive sound waves, with low frequency, whereas the optical modes are vibrations of atoms or molecules relative to one another, typically having a higher frequency. In the absence of low-lying traverse optical (rattling) and dispersionless (single frequency) Einstein modes, the heat capacity expression can be written as:

$$C_p(T) = \gamma T + [9s_D R \left(\frac{T}{\theta_D}\right)^3 \int_0^{\theta_D/T} \frac{x^4}{(e^x-1)(1-e^{-x})}\, dx] \quad (6)$$

where $\gamma$ is the Sommerfeld coefficient related to electronic contribution to the heat capacity, $\Theta_D$ is the Debye temperature associated with maximum or cutoff frequency for lattice vibrations, $s_D$ refers to the total number of atoms in the unit cell, and $R$ is the universal gas constant. We observed no characteristic $T_{max}$ peak in the $C_p/T^3$ vs. log($T$) plot, validating the

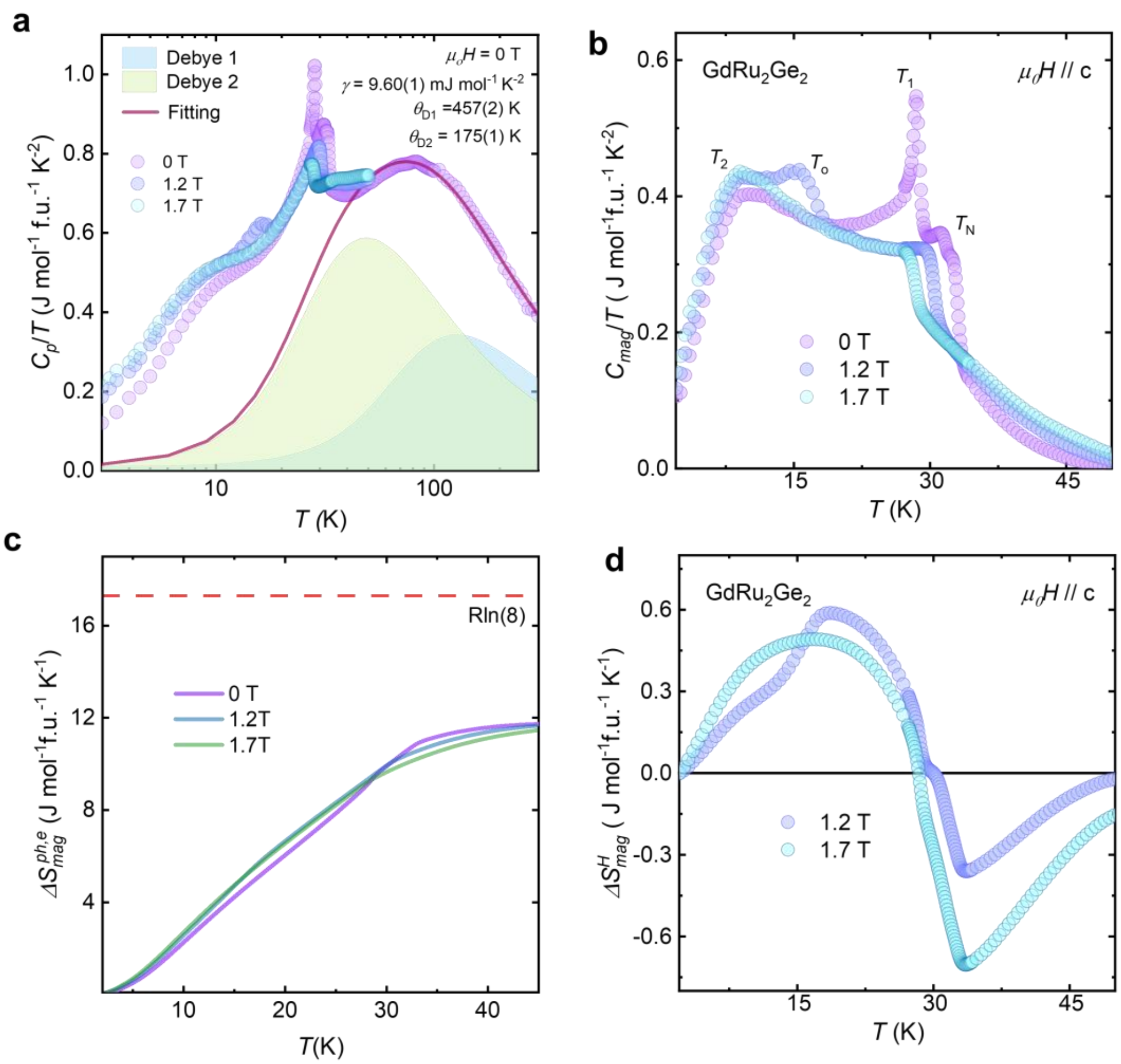

**Figure 5.** (**a**) Molar heat capacity over temperature ($C_p/T$) vs. temperature at different fields. The dark-red line shows the calculated phonons using the two-Debye model for molar heat capacity at $\mu_0H$ = 0 T. (**b**) Magnetic contribution to the molar heat capacity over temperature ($C_{mag}/T$) vs. temperature at different fields, (**c**) Magnetic entropy change w.r.t phonon and electronic contribution, $\Delta S_{mag}^{ph,e}$, and (**d**) Magnetic entropy change w.r.t field, $\Delta S_{mag}^{H}$. The expected value for Rln(8) is indicated in dashed-red line.

absence of Einstein (optic) modes (**Figure S5**). We attempted to estimate the low-temperature $\gamma$ value and $\beta$ value using the $C_p/T$ vs $T^2$ linear fitting ($C_p = \gamma T + \beta T^3$), but we observed a linear region at 2 K ≤ $T$ ≤ 5 K due to the low-lying electronic Schottky anomaly ($T_2$) associated with $Gd^{3+}$ magnetic ion ground state splitting at 5 K ≤ $T$ ≤ 12 K. Such a narrow range fit may not sufficiently reflect the $\gamma$ and $\beta$ values and makes magnetic contribution estimation challenging. Therefore, we followed the below-mentioned method to estimate the electronic and phonon (lattice vibrations) contributions to heat capacity. At high temperatures, the heat capacity saturates to 120 J mol$^{-1}$ K$^{-1}$, indicating the high temperature limiting value corresponding to 5 atoms satisfied according to the Dulong Petit law ($3nR$ =

124.71 J mol$^{-1}$ K$^{-1}$). Using the following model, we have chosen a temperature range 50 K ≤ $T$ ≤ 300 K to fit the heat capacity. The chosen model includes two Debye (acoustic) modes and $\gamma$ electronic contribution (Eq. 7-8).

$$\frac{C_p}{T} = \frac{C_{Debye\,(1)}}{T} + \frac{C_{Deybe(2)}}{T} + \gamma \qquad (7)$$

$$C_{Debye} = 9Rs_D \left(\frac{T}{\theta_D}\right)^3 D(\frac{\theta_D}{T}) \qquad (8)$$

where $s_D$ is the number of oscillators per formula unit and is generally connected to the total number of atoms in the unit cell. The two Debye functions are warranted to account for low-temperature deviation from quadratic frequency variation of the phonon density of states. Initially, we fixed the $\gamma$ value based on the estimate from the density of states through the band structure calculation ($\eta(\mathcal{E}_F)$ = 2.5 states/eV, and $\gamma$ = 5.9 mJ mol$^{-1}$ K$^{-1}$). During the fitting process, we allowed only the two Debye temperatures to vary and attempted several pairings of $s_{D1}$ and $s_{D2}$ values by keeping the total number of oscillators to 5. We found that the model having $s_{D1}$ = 3 and $s_{D2}$ = 2 explains the experimental heat capacity data well. Once the phonon contribution to the heat capacity was well-fitted, the $\gamma$ value was allowed to vary, and the fitting value is shown in **Figure 5a**. We then simulated the electronic and phonon contributions to the heat capacity up to $T$ = 2 K using the estimated $\theta_{D1}$, $\theta_{D2}$, and $\gamma$ values. The electronic contribution $\gamma$ is 9.60(1) mJ mol$^{-1}$ K$^{-2}$, which is in the same order of magnitude as other magnetic metals.[24] The increased $\gamma$, compared to the theoretical value from the band structure calculation, can stem from a number of possible electron interactions, such as electron-electron, electron-phonon, electron-magnon, and spin fluctuations. The magnetic contribution to the heat capacity was estimated at 2 K ≤ $T$ ≤ 50 K by subtracting the simulated heat capacity. The temperature dependence of $C_{mag}$ is shown in **Figure 5b**. Anomalies in heat capacity data represent the bulk nature of any phase transitions; however, both temperature-driven and entropy-driven transitions show a positive peak in heat capacity measurement. To determine whether the heat capacity peak associated with the skyrmion phase ($T_0$) corresponds to positive magnetoentropy, we estimated $\Delta S^{ph,e}_{mag}$ as a function of temperature for all fields using Equation 9.

$$\Delta S^{ph,e}_{mag} = \int_0^T \frac{C_{mag}}{T}\, dT \qquad (9)$$

The subscript denotes the magnetic contribution to the heat capacity, while the superscript indicates the estimation of magnetic entropy relative to the electronic and phonon contributions to the heat capacity. The maximum entropy recovered at $T$ = 50 K is ~12 J mol$^{-1}$ f.u.$^{-1}$ K$^{-1}$, which is lower than expected for $Gd^{3+}$ local moment ions (**Figure 4c**). A plausible reason for this underestimation can be attributed to the presence of a Schottky-type hump, as we mentioned previously. However, our goal here is to make a qualitative analysis of the peak shape corresponding to the skyrmion phase and not to quantify the isothermal entropy change.

Positive entropy change associated with skyrmion formation—an insight well established in the literature—highlights that skyrmion is a high-entropy phase. There are many contributions to heat capacity data, such as phonon and electronic effects, making it difficult to isolate subtle magnetic entropy changes. By carefully modeling the phonon and electronic background and extracting the entropy $S$ ($T$, $H$), we were able to resolve these fine features. Therefore, the observed positive entropy is consistent with the entropy-driven skyrmion phase transition, as governed by free energy considerations. We used $\Delta S_{mag}(0, T)$ as a reference to estimate $\Delta S_{mag}^{H} = \Delta S(H,T) - \Delta S(0,T)$. The magnetic entropy estimated shows a positive peak at $T$ = 20 K and a negative peak at $T$ = 33 K corresponding to the origin of the skyrmion phase and AFM ordering, respectively (**Figure 5d**). We illustrated that positive entropy mapping from heat capacity analysis complements insights into skyrmions evolution gained from magnetization measurements. Furthermore, this thermodynamic signature helps to differentiate the skyrmion phase from other magnetic phase transitions that are topologically trivial. Even though magnetization measurements are more sensitive, heat capacity probes the bulk nature of phase transitions. This combined analysis highlights that the skyrmion formation in $GdRu_2Ge_2$ is a lattice phenomenon with periodic modulation, not an isolated pocket or a bubble randomly forming in the lattice, agreeing well with the reported literature.[10]

To investigate the evidence for the topological features of $GdRu_2Ge_2$, charge transport properties were measured under magnetic fields down to $T$ = 2 K (**Figure 6**). The zero-field temperature dependence of resistance exhibits a typical linear behavior at 50 ≤ $T$ ≤ 300 K, as expected for a metallic sample. At $T$ < 33 K, we observed a rapid decrease in resistivity, indicating the suppression of spin-disorder scattering, which is expected in all magnetically ordered systems. The resistivity at $T$ = 2 K was measured to be 2 μΩ-cm, and the residual resistivity ratio (RRR) was found to be ~ 17. This relatively large RRR value confirms the high quality of the sample. The temperature dependence of $\rho_{xx}$ for several applied fields is shown in **Figure 6a**. As the field increases, the magnetic ordering temperature decreases. When the field is increased to 2 T, the magnetic ordering becomes smeared out. In applied fields, we would typically expect a decrease in magnetic ordering temperature for a conventional AFM and smearing out of magnetic ordering for an FM system.[25] Since we observed both the temperature reduction and smearing of the magnetic transitions in response to the applied field for $GdRu_2Ge_2$, this indicates the presence of competing magnetic interactions. The magnetoresistance ($MR$) is defined as:

$$MR = \frac{R(H) - R(0)}{R(0)} \% \qquad (10)$$

where $R(H)$ and $R(0)$ are resistance values at an applied field $\mu_0 H$ and zero field, respectively.

We observed a large negative *MR* with a peak-like behavior centered around the magnetic transition (**Figure S6**). The low-temperature positive *MR* arises from the dominance of the Lorentz contribution. This is owing to the reduction in the spin-wave-related *MR* upon the magnetic saturation. Apart from the conventional negative *MR* expected for magnetic metals, we do not observe any signatures of topological features in the *MR*, as carrier scattering along the longitudinal direction is often insensitive to nontrivial band topology or scalar spin chirality.

To understand these effects, we carried out Hall resistivity ($\rho_{xy}$) measurements. **Figure 6b** shows the temperature variation of $\rho_{xy}$ at 2 K ≤ $T$ ≤ 50 K and different applied fields. We observed negative $\rho_{xy}$ in the entire temperature and field range, indicating that the majority of charge carriers are holes. However, the high-field $\rho_{xy}$ data exhibit a nonlinear behavior, which could indicate a multiband effect. The $\rho_{xy}$ vs. $T$ plot (**Figure 6b**) shows a negative hump-like feature near the magnetic transition at lower fields $\mu_0 H$ = 0.5, 1 and 1.2 T, and a positive hump at higher fields $\mu_0 H$ = 1.5 and 2 T. This behavior suggests subtle variations in the electronic structure where charge carriers passing through the magnetic spin swirls (skyrmions) could experience an emergent field, known as the Berry curvature.

In general, the total Hall resistivity $\rho_{xy}$ can be expressed as:

$$\rho_{xy} = \rho_{xy}^{NHE} + \rho_{xy}^{AHE} + \rho_{xy}^{THE} \qquad (11)$$

where the first term corresponds to the normal (or ordinary) Hall effect (NHE), which arises due to the Lorentz force and typically varies linearly with the applied magnetic field. The second term represents the anomalous Hall effect (AHE), which originates from anomalous scattering mechanisms in magnetically ordered materials. This contribution is influenced by side-jump and/or skew scattering of charge carriers due to magnetization. The third term accounts for the topological Hall effect (THE), which appears only in materials with nontrivial electronic states exhibiting Berry curvature in momentum space or in materials hosting skyrmions or magnetic spin vortices in the real-space lattice[2d, 26]. The topological Hall contribution can be estimated by carefully removing the $\rho_{xy}^{NHE}$ and $\rho_{xy}^{AHE}$.

$$\Delta\rho_{xy} \text{ or } \rho_{xy}^{THE} = \rho_{xy} - (\rho_{xy}^{NHE} + \rho_{xy}^{AHE}) \qquad (12)$$

To extract Hall coefficients ($R_0$), linear Hall fittings were performed at $T$ = 100 K, 200 K, and 300 K using the normal Hall expression $\rho_{xy} = R_0 H$. The extracted $R_0$ values are positive at $T$ = 100 K and 200 K, indicating that hole-like carriers dominate the conduction in this temperature range. However, at $T$ = 300 K, the Hall resistivity displays two distinct linear regimes: a low-field fit (0 – 5 T) and a high-field fit (5 – 7 T), both of which yield negative $R_0$ values, suggesting a crossover to electron-like conduction or a possible multiband transport scenario at high temperatures. The corresponding carrier densities range from 0.24 x $10^{23}$ to 13.2 × $10^{23}$ $cm^{-3}$, consistent with a metallic behavior (**Table S5**).

At low temperatures, the anomalous Hall contribution was estimated using two widely followed methods.[27] In the first method, the total Hall contribution is written as the following equation:

$$\rho_{xy} = \mu_0 R_0 H + S_H \rho_{xx}^2 M + \rho_{xy}^T \qquad (13)$$

where $\rho_{xx}$ is longitudinal resistivity and $S_H$ is the anomalous Hall scaling factor. $S_H$ is extracted from the slope of the $\frac{\rho_{xy}}{\mu_0 H} vs\, \rho_{xx}^2 M$ plot above the critical field $H_c$ where THE contribution vanishes.[27c] In the second method, the anomalous Hall resistivity is determined by analyzing high-field data, where the magnetization saturates, and the THE is absent. For systems such as NiMnIn and NiMnGa,[28] the AHE was extracted by plotting $\frac{\rho_{xy}}{\mu_0 H}$ versus $\frac{M}{\mu_0 H}$ leading to a linear relation:

$$\frac{\rho_{xy}}{\mu_0 H} = R_0 + R_s \cdot \frac{M}{\mu_0 H} \qquad (14)$$

where the intercept corresponds to the ordinary Hall coefficient $R_0$, and the slope gives the anomalous Hall coefficient $R_s$. In our analysis, we adopted a similar approach by modeling the Hall resistivity as:

$$\rho_{xy}(H) = R_0 H + R_s M(H) \qquad (15)$$

$$\rho_{xy}^{THE}(H) = \rho_{xy}^{\text{meas}}(H) - \big(R_0 H + R_s M(H)\big) \qquad (16)$$

where the first term represents the normal Hall effect and the second term accounts for the anomalous Hall contribution in equation 15. The fit was performed in the high-field region (5.5 - 7 T) using equation 15, where the Hall resistivity is saturated and the THE is expected to vanish. The extracted coefficients were then used to reconstruct the background over the full field range, and the topological Hall resistivity $\rho_{xy}^T$ was obtained by subtracting the fitted background from the measured Hall data according to equation 16. This method is valid under the condition that magnetoresistance is negligible or small in the fitting region. In our case, the field variation of $\rho_{xx}$ below 25 K is negligible, and the negative MR observed between 25 K and 33 K is attributed to spin-wave or spin-disorder scattering, which does not extend below 25 K. Therefore, we can reasonably justify that $\rho_{xx}$ is nearly constant as a function of applied field in the low-temperature regime (**Figure 6a**).

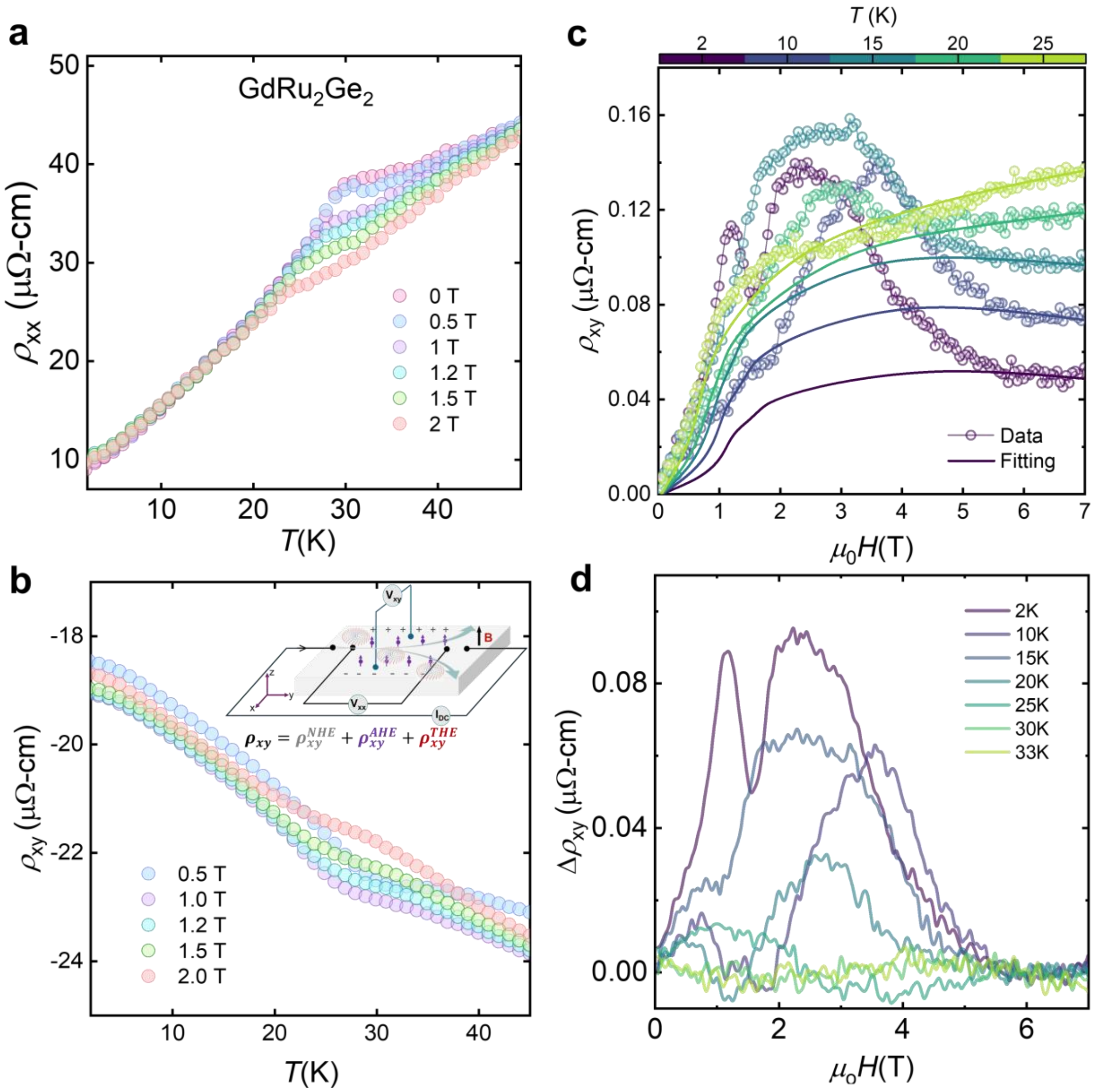

**Figure 6**. Transport properties of $GdRu_2Ge_2$ as a function of temperature and magnetic field. (**a**) Longitudinal resistivity ($\rho_{xx}$) as a function of temperature ($T$) for different applied magnetic fields ($\mu_0 H$), (**b**) Transverse (Hall) resistivity ($\rho_{xy}$) as a function of $T$ for different field values, inset showing a schematic illustration of measurement geometry for the linear four probe resistivity and Hall resistivity, (**c**) Hall resistivity ($\rho_{xy}$) and fitting showing normal Hall effect and anomalous Hall effect as a function of $\mu_0 H$ under different $T$, and (**d**) Topological Hall resistivity ($\Delta\rho_{xy}$), obtained after subtracting normal and anomalous Hall contributions. The oscillatory behavior suggests a nontrivial magnetic texture.

**Figure 6c** shows the temperature dependence of the total Hall resistivity, along with the model fit using Equation 13. We observe that the saturation value of the Hall resistivity decreases as the temperature decreases, whereas the magnetization continues to increase. This indicates a significant suppression of the anomalous Hall contribution at low temperatures, as reflected in the drastic reduction of the fitted $R_s$ values (**Table S5**). At 30 K and 33 K, the linear behavior at high fields allows for a reliable estimation of the ordinary Hall coefficient $R_0$. In contrast, the low-temperature data exhibit slight oscillations, which may stem from limitations in the fitting model. Nonetheless, the extracted values are of the same order of magnitude as those reported in known skyrmion compounds. **Figure 6d** presents the temperature variation of the topological Hall resistivity, which exhibits a distinct double-peak structure at $T$ = 2 to 20 K, which can be related to the two skyrmion regions observed. The topological features observed in Hall effect measurements support the observation from magnetization and heat capacity analysis.

We estimated the effective magnetic field $B_{eff}$, arising from the noncoplanar spin texture using the relation [26d, 27c]:

$$B_{\text{eff}} = -\frac{h}{e}\left(\frac{\sqrt{3}}{2\lambda_s^2}\right) \qquad (17)$$

where $\lambda_s$ is the helical period associated with the magnetic ordering wavevector $\boldsymbol{q} = (0.22, 0, 0)$ in $GdRu_2Ge_2$. Taking the lattice constant $a = 4.2$ Å, we estimate $\lambda_s = \frac{a}{0.22} \approx 1.91$ nm.[10] The substitution of this value into the above expression yields an expected effective magnetic field of $B_{\text{eff}} \approx -981.5\ T$, which is two orders more than values reported for non-centrosymmetric based skyrmions such as MnSi and FeGe.[26d, 27c] The emergent magnetic field arising from the Berry curvature is a fictitious field associated with the characteristic length scale of the skyrmion size and should not be directly compared to the magnitude of an externally applied magnetic field.

The topological Hall resistivity is estimated using $\rho_{xy}^{\text{THE}} = PR_0B_{\text{eff}}$, where $P$ = −0.14 is the spin polarization obtained from the density of states analysis and $R_0 = 2.35$ n Ω cm/T is the ordinary Hall coefficient obtained from our 2 K data. This yields a value of $\rho_{xy}^{\text{THE, est}} = -0.315$ μΩ cm, comparable to the experimentally observed maximum of $\rho_{xy}^{\text{THE, exp}} = 0.102$ μΩ cm. This small numerical discrepancy in both magnitude and sign suggests additional sources such as chiral fluctuations or multi-band effects.

To get insight into how the electronic structure of $GdRu_2X_2$ determines their physical properties, pseudopotential spin-polarized DFT calculations were performed using the Quantum Espresso Software package[13] employing a projected augmented wave (PAW) method. The results clearly demonstrate some common electronic structure features of the $ThCr_2Si_2$ structure type (**Figure 7**). The spins of the Gd-4*f* states are polarized, which then

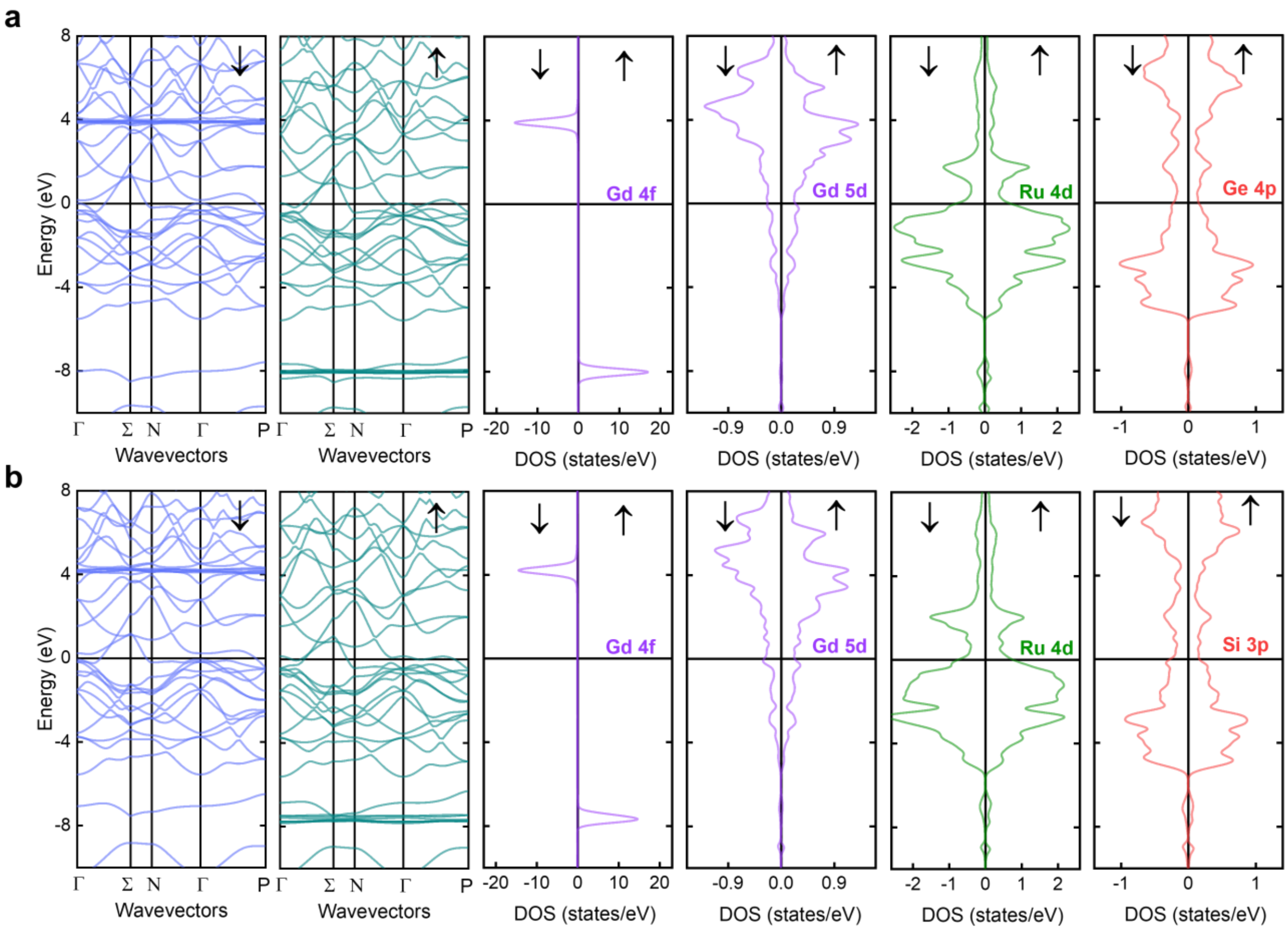

**Figure 7**. Spin-polarized band structures showing bands around the Fermi level and spin-polarized DOS of (**a**) $GdRu_2Ge_2$, and (**b**) $GdRu_2Si_2$

polarizes the Ru-4*d* and X-3*p* states. The contribution of Gd-4*f* states is localized, deep in low energy ~-8 eV for majority spins and slightly above the Fermi level ($E_F$) energy ~ 4 eV for minority spins. The spin-polarized band structure and density of states (DOS) of $GdRu_2X_2$ display a metallic behavior, where multiple bands cross $E_F$ and finite DOS at $E_F$. These features prove that interactions between the localized Gd-4*f* magnetic moments are mediated through the itinerant electrons—RKKY interactions. Comparing the DOS and band structures in **Figure 7**, similar features are represented within the isostructural skyrmion system $GdRu_2X_2$ (**Figure 7b**). The energy dispersion of DOS in $GdRu_2Ge_2$ is more dispersed than that in $GdRu_2Si_2$ due to more diffused Ge-4*p* vs. Si-3*p* (**Figure 7 and Figure S11**). DOS itself is not the only proof of the extended orbital features, and rigorous chemical bonding analysis is needed to provide a more complete argument for the contribution of Ge-4*p* vs. Si-3*p* (**Tables S6** and **S7**). Additionally, chemical bonding information derived from crystal orbital Hamilton population (COHP) and integrated ICOHP provides insights into the strength of the orbital overlap along Gd-Ru-Gd ($J_1$) and Gd-X-Gd ($J_2$) pathways (**Table S6**).

Increased ICOHP values observed for Gd-X bonding indicate a stronger overlap in the $J_2$ pathway, supporting the exchange interactions $J_2 > J_1$.

Taken together, our exchange-correlation calculations and magnetic and transport property analyses enabled us to establish a connection between the competing exchange interactions and the skyrmion phase transition conditions (**Table 2**).

**Table 2**. Summary of the critical field and temperature conditions at which skyrmions evolve and the nearest neighbor exchange interaction $J_1$ in $GdRu_2X_2$ (A = Si, Ge).

| $GdRu_2X_2$ | Critical field (T) | Critical Temperature (K) | $J_1$ (meV) |
|---|---|---|---|
| $GdRu_2Si_2$ | $2 \leq \mu_0 H \leq 2.5$ | $2 \leq T \leq 20$ | -0.06 (2) |
| $GdRu_2Ge_2$ | $0.9 \leq \mu_0 H \leq 1.2$<br>$1.3 \leq \mu_0 H \leq 1.7$ | $2 \leq T \leq 30$ | -6.12 (2) |

The results highlight that increased competing exchange interactions at similar energy scales within the Gd square lattice stabilize multiple topological spin states while making the phase transition to skyrmions more accessible at higher temperatures and lower fields. We propose that substituting Si-3*p* with more extended Ge-4*p* orbitals enhances orbital overlap, thus increasing the nearest neighbor exchange interaction $J_1$ in $GdRu_2Ge_2$. Similar examples include $EuAl_4$, $EuGa_2Al_2$, and $EuGa_4$. While these Eu-based materials are isostructural tetragonal centrosymmetric systems, they display skyrmions at different critical fields and temperatures 0.75 T ≤ $\mu_0 H$ ≤ 1.4 T; 2 K ≤ $T$ ≤ 13 K, 1.2 T ≤ $\mu_0 H$ ≤ 1.6 T; 2 K ≤ $T$ ≤ 9 K, 4 T ≤ $\mu_0 H$ ≤ 7 T; 2 K ≤ $T$ ≤ 8 K respectively.[4d, 6a, 6b] This can be attributed to the chemically enabled tunability of their competing exchange interactions and, thus, their skyrmion phase transition conditions. This approach can be extended beyond centrosymmetric systems. MnSi and MnGe are isostructural noncentrosymmetric chiral magnets that host skyrmions at 0.1 T ≤ $\mu_0 H$ ≤ 0.25 T; 26 K ≤ $T$ ≤ 30 K, 0 T ≤ $\mu_0 H$ ≤ 2.4 T; 2 K ≤ $T$ ≤ 170 K and respectively.[3i, 27c, 29]

## CONCLUSIONS

Developing skyrmion materials for spintronics requires the tunability and control of critical temperature and magnetic field conditions at which skyrmions emerge. Our results demonstrate a new, fundamental framework for understanding and realizing rich topological spin phases. We investigated the impact of competing exchange interactions on the skyrmion evolution in $GdRu_2X_2$ (X = Si, Ge) and supplemented with crystal and electronic structure, magnetic, thermodynamic, and transport properties of the $GdRu_2Ge_2$. The $J$ coupling constants, obtained from our DFT calculations, reveal appreciably stronger FM-AFM competing exchange interactions in $GdRu_2Ge_2$ compared to $GdRu_2Si_2$. This is attributable to the more extended Ge-4$p$ orbitals than the Si-3$p$ orbitals, thus improving the overlap of the atomic interacting wave functions in $GdRu_2Ge_2$. The enhanced FM-AFM competing interactions in the Ge compound facilitate the phase transition to skyrmions at higher temperatures and lower fields. $GdRu_2Ge_2$ displays two skyrmion phases at 0.9 T ≤ $\mu_0H$ ≤ 1.2 T and 1.3 T ≤ $\mu_0H$ ≤ 1.7 T, 2 K ≤ $T$ ≤ 30 K, lower field and higher temperature than that in $GdRu_2Si_2$ (2 T ≤ $\mu_0H$ ≤ 2.5 T, 2 K ≤ $T$ ≤ 20 K).[8a, 9] The skyrmion evolution in $GdRu_2Ge_2$ is proven through magnetization, magnetoentropic mapping, heat capacity, transport, and Hall effect measurements. Further studies are underway to directly visualize skyrmions formation and experimentally determine magnetic exchange interactions. Our work provides a significant step forward for atomically enabled tunability of competing exchange interactions in magnetic metals while linking this insight to the critical conditions at which skyrmions form. This research approach can be extended beyond centrosymmetric magnets for developing new skyrmions with designer working conditions for spintronics and logic constructs.

**ASSOCIATED CONTENT**

Experimental section

**Synthesis of polycrystalline $GdRu_2Ge_2$.**

$GdRu_2Ge_2$ was synthesized by arc-melting high-purity elements (purity > 99.9 wt %) using a Centorr vacuum arc-melter under a continuous flow of high-purity argon to prevent oxidation. To ensure homogeneity, the ingot was flipped and remelted several times. After arc-melting, the sample was sealed in an evacuated quartz tube and annealed at 800 °C for one week. No significant differences were observed in the structural parameters or physical properties between the annealed and unannealed samples. Rod-shaped, small, single crystals were picked from the smashed metal ingot.

**Synthesis of $GdRu_2Ge_2$ single crystals.**

To grow sizable crystals tilted LASER diode floating zone (Tilted LDFZ) furnace was employed. Approximately ~0.7 g of melted $GdRu_2Ge_2$ button was connected to the Ru wire by melting them together (feed rod). This was then mounted onto the Alumina sample holder. Feed rod and seed rod (another Ru wire) were attached to the lower and upper shaft, respectively of the Tilted LDFZ with IR lasers (974 nm). Both upper and lower shafts were adjusted for the proper axial alignment. Laser positions were adjusted to at $3^0$ tilt angle. First, the top of the feed rod was melted and connected to the seed to create the molten zone. During this, both seed and feed rods were rotated clockwise at a rate of 7.6 and 7.4 rpm/hr. Once a stable, sizable molten zone was generated, rods were counter-rotated at the same speeds. During the growth, feed and seed rods were translated upwards with 10 and 3, 8 and 3, and 6 and 4 mm/hr, respectively. The inert atmosphere was maintained by flowing Ar gas flow at a rate of 0.75 L/min. Grown crystal boule was characterized using the powder X-ray and Laue diffractions. Multiwire real time back-reflection X-ray Laue diffractometer (MWL120) with the X-ray beam of about 1mm in diameter, 7.2 kV, 10 mA was utilized to check the crystalline quality and cut to the required dimensions using a diamond saw.

**Single Crystal X-ray diffraction.**

A single metallic black block-shaped crystal of $GdRu_2Ge_2$ with dimensions $0.12 \times 0.10 \times 0.06$ mm$^3$ was used. It was mounted on a loop with paratone on a Bruker D8 Quest diffractometer. The crystal was kept at a constant $T$ = 300 K during data collection. Data were measured using $\omega$ scans with Mo $K_a$ radiation ($\lambda$ = 0.71073 Å) and a Photon 3 detector. Data processing (SAINT) and scaling (SADABS) were performed using the Apex4 software system. The structure was solved by intrinsic phasing (SHELXT) and refined by full-matrix least-squares minimization on $F^2$ (SHELXL) using the SHELXTL software program.[30] All atoms were refined anisotropically.

**Powder X-ray diffraction.**

Powder X-ray diffraction (PXRD) measurements were performed using Rigaku Ultima IV diffractometer equipped with Cu Kα radiation ($\lambda$ = 1.5406 Å). Data were collected in the 2 θ range of 5−95° at 0.2°/min. Rietveld refinement of the XRD pattern was performed using TOPAS Academic V6.

**Magnetization and Specific Heat.**

DC magnetization measurements on $GdRu_2Ge_2$ were performed with the Vibrating sample magnetometer (VSM) option of Quantum Design Physical Properties Measurement System (PPMS) between 2 K ≤ $T$ ≤ 300 K at 0 T ≤ $\mu_0 H$ ≤ 7 T. Heat capacity was measured using the PPMS, employing the semi adiabatic pulse technique from $T$ = 2 to 300 K.

**Transport measurements.**

The electrical and magneto-transport properties (resistivity, magnetoresistance, and Hall effect) were measured in the temperature range of 2–300 K and up to a magnetic field of 5 T using the DC transport option of a Physical Property Measurement System (PPMS, Quantum Design Inc., San Diego) with the conventional four-probe method. Gold leads were attached to the sample using DuPont 4922N silver paint and mixed with butyl acetate for better adhesion. For resistivity and magnetoresistance (MR) measurements, we used a bar-shaped polycrystalline sample with dimensions $2 \times 0.6 \times 0.5$ mm$^3$, while a pressed polycrystalline pellet with dimensions $2 \times 1.5 \times 0.05$ mm$^3$ was used for Hall effect measurements. This approach was chosen to satisfy the $w \gg t$ geometry required for observing an appreciable Hall signal, as the aspect ratio $w/t \approx 20$ is more suitable for Hall measurements in polycrystalline samples. For Hall measurements, the magnetic field was applied perpendicular to the flat surface of the pellet, corresponding to a nominal field direction perpendicular to the current path. Symmetric and antisymmetric components of the Hall signal were separated by performing field-reversal measurements. The values of $\rho_{xy}$ reported in this work correspond to the antisymmetric part of the Hall voltage, extracted as $\rho_{xy} = \frac{\rho_{xy}(+B)-\rho_{xy}(-B)}{2}$. This antisymmetrization procedure ensures the removal of longitudinal voltage contributions due to misalignment of voltage contacts.

**Density Functional theory calculations.**

Pseudopotential band structure and polarized density of states were calculated using the pw.x program in the Quantum Espresso (QE) software package,[13] with the Generalized Gradient Approximation of the exchange-correlation potential and Hubbard U correction (GGA+U)for the strong correlations in 4*f* bands (6.7 eV for Gd-4*f*) [31] of the exchange-correlation potential with the PBEsol parametrization.[32] Projector-augmented wave (PAW) potentials for Ge, and Ru were taken from the PSlibrary v.1.0.0 set, and for Gd, PAW potential developed by VLab was used.[14d, 33] The self-consistencies were carried out using 13 ×13 × 6 *k*-mesh in the irreducible Brillouin zone. Kinetic energy cutoff for charge density and wavefunctions was set to 60 Ry and 412 Ry. The pseudo potential DFT calculations for the *J*-coupling constant used the same parameter, except the k-mesh was changed to 3 × 3 × 2 to account for the (2*a*,2*b*,*c*) super cell.

**Supporting Information**. Additional data analysis, tables, figures, including crystallographic data, XRD, SEM-EDS, magnetization, heat capacity, transport measurements TGA, DOS.

**AUTHOR INFORMATION**

**Corresponding Author**

*Thao T. Tran, email: thao@clemson.edu.

**Author Contributions**

The manuscript was written through contributions of all authors. All authors have given approval to the final version of the manuscript.

**Funding Sources**

The work at Clemson University was supported by the National Science Foundation under CAREER Award NSF-DMR-2338014. X.H., R.K., and T.T.T thank the Arnold and Mabel Beckman Foundation for a 2023 BYI award to T.T.T. Research performed at Gdansk Tech was supported by the National Science Center (Poland) OPUS grant no. UMO-2022/45/B/ST5/03916.

**ACKNOWLEDGMENTS**

We thank Dr. C. McMillen, Dr. R. Sachdeva, and Dr. Matthew Powell for their assistance in X-ray diffraction, TGA, and some magnetization, heat capacity, and transport measurements, respectively.

# Supporting Information

## Atomically Modulating Competing Exchange Interactions in Centrosymmetric Skyrmion Hosts $GdRu_2X_2$ (X = Si, Ge)

Dasuni N. Rathnaweera,[1] Xudong Huai,[1] K. Ramesh Kumar,[1] Michał J. Winiarski,[2] Tomasz Klimczuk,[2] Allana G. Iwanicki,[3,4] Satya Kushwaha,[3,4] Martin Mourigal,[5] Tyrel M. McQueen,[3,4,6] and Thao T. Tran[1,*]

[1]Department of Chemistry, Clemson University, Clemson, South Carolina, 29634, USA

[2]Faculty of Applied Physics and Mathematics and Advanced Materials Center, Gdansk University of Technology, ul. Narutowicza 11/12, 80-233 Gdansk, Poland

[3]Institute for Quantum Matter, William H. Miller III Department of Physics and Astronomy, Johns Hopkins University, 3400 N. Charles Street, Baltimore, MD 21218, USA

[4]Department of Chemistry, Johns Hopkins University, 3400 N. Charles Street, Baltimore, MD, 21218, USA

[5]School of Physics, Georgia Institute of Technology, Atlanta, GA, 30332, USA

[6]Department of Materials Science and Engineering, Johns Hopkins University, Baltimore, 21218, MD, USA

**Experimental section**

**List of Figures:**

**List of Tables:**

**Table S5.** Transverse resistivity fitting parameters and error bar plot.

**Table S6.** Integrated crystal orbital Hamilton population (ICOHP) values indicating the bond strength

**Table S7**. Integrated crystal orbital overlap (ICOOP) values indicating the orbital overlap.

Experimental section

**Synthesis of polycrystalline $GdRu_2Ge_2$.**

$GdRu_2Ge_2$ was synthesized by arc-melting high-purity elements (purity > 99.9 wt %) using a Centorr vacuum arc-melter under a continuous flow of high-purity argon to prevent oxidation. To ensure homogeneity, the ingot was flipped and remelted several times. After arc-melting, the sample was sealed in an evacuated quartz tube and annealed at 800 °C for one week. No significant differences were observed in the structural parameters or physical properties between the annealed and unannealed samples. Rod-shaped, small, single crystals were picked from the smashed metal ingot.

**Synthesis of $GdRu_2Ge_2$ single crystals.**

Approximately ~0.7 g of melted $GdRu_2Ge_2$ button was connected to the Ru wire by melting them together (feed rod). This was then mounted onto the Alumina sample holder. Feed rod and seed rod (another Ru wire) were attached to the lower and upper shaft, respectively of the tilted LASER diode floating zone with IR lasers (974 nm). Both upper and lower shafts were adjusted for the proper axial alignment. Laser positions were adjusted to at $3^0$ tilt angle. First, the top of the feed rod was melted and connected to the seed to create the molten zone. During this, both seed and feed rods were rotated clockwise at a rate of 7.6 and 7.4 rpm/hr. Once a stable, sizable molten zone was generated, rods were counter-rotated at the same speeds. During the growth, feed and seed rods were translated upwards with 10 and 3, 8 and 3, and 6 and 4 mm/hr, respectively. The inert atmosphere was maintained by flowing Ar gas flow at a rate of 0.75 L/min. Grown crystal boule was characterized using the powder X-ray and Laue diffractions. Multiwire real time back-reflection X-ray Laue diffractometer (MWL120) with the X-ray beam of about 1mm in diameter, 7.2 kV, 10 mA was utilized to check the crystalline quality and cut to the required dimensions using a diamond saw.

**Single Crystal X-ray diffraction.**

A single metallic black block-shaped crystal of $GdRu_2Ge_2$ with dimensions 0.12 × 0.10 × 0.06 $mm^3$ was used. It was mounted on a loop with paratone on a Bruker D8 Quest diffractometer. The crystal was kept at a constant $T$ = 300 K during data collection. Data were measured using $\omega$ scans with Mo $K_\alpha$ radiation (λ = 0.71073 Å) and a Photon 3 detector. Data processing (SAINT) and scaling (SADABS) were performed using the Apex4 software system. The structure was solved by intrinsic phasing (SHELXT) and refined by full-matrix least-squares minimization on $F^2$ (SHELXL) using the SHELXTL software program.[1] All atoms were refined anisotropically.

**Powder X-ray diffraction.**

Powder X-ray diffraction (PXRD) measurements were performed using Rigaku Ultima IV diffractometer equipped with Cu Kα radiation (λ = 1.5406 Å). Data were collected in the 2 θ range of 5−95° at 0.2°/min. Rietveld refinement of the XRD pattern was performed using TOPAS Academic V6.

**Magnetization and Specific Heat.**

DC magnetization measurements on $GdRu_2Ge_2$ were performed with the Vibrating sample magnetometer (VSM) option of Quantum Design Physical Properties Measurement System (PPMS) between 2 K ≤ $T$ ≤ 300 K at 0 T ≤ $\mu_0 H$ ≤ 7 T. Heat capacity was measured using the PPMS, employing the semiadiabatic pulse technique from $T$ = 2 to 300 K.

**Transport measurements.**

The electrical and magneto-transport properties (resistivity, magnetoresistance, and Hall effect) were measured in the temperature range of 2–300 K and up to a magnetic field of 5 T using the DC transport option of a Physical Property Measurement System (PPMS, Quantum Design Inc., San Diego) with the conventional four-probe method. Gold leads were attached to the sample using DuPont 4922N silver paint and mixed with butyl acetate for better adhesion. For resistivity and magnetoresistance (MR) measurements, we used a bar-shaped polycrystalline sample with dimensions 2 × 0.6× 0.5 $mm^3$, while a pressed polycrystalline pellet with dimensions 2 × 1.5 × 0.05 $mm^3$ was used for Hall effect measurements. This approach was chosen to satisfy the $w \gg t$ geometry required for observing an appreciable Hall signal, as the aspect ratio $w/t \approx 20$ is more suitable for Hall measurements in polycrystalline samples. For Hall measurements, the magnetic field was applied perpendicular to the flat surface of the pellet, corresponding to a nominal field direction perpendicular to the current path. Symmetric and antisymmetric components of the Hall signal were separated by performing field-reversal measurements. The values of $\rho_{xy}$ reported in this work correspond to the antisymmetric part of the Hall voltage, extracted as $\rho_{xy} = \frac{\rho_{xy}(+B) - \rho_{xy}(-B)}{2}$. This antisymmetrization procedure ensures the removal of longitudinal voltage contributions due to misalignment of voltage contacts.

**Thermogravimetric analysis.**

Thermogravimetric analysis and differential scanning calorimetry measurements were performed using a TA SDTQ600 Instrument. Approximately 15 mg of $GdRu_2Ge_2$ powder was placed in an alumina crucible and heated at a rate of 20 °C/min from room temperature to 1000 °C under flowing nitrogen of flow rate: 100mL/min; Figure S1).

**Scanning electron microscopy analysis.**

SEM-EDS analysis was conducted using a Hitachi SU5000 VP-SEM equipped with a Schottky Field Emission source for imaging. An Oxford EDS system was utilized to verify the Gd:Ru:Ge ratio, and the data was processed using AZtecLive software.

**Density Functional theory calculations.**

Pseudopotential band structure and polarized density of states were calculated using the pw.x program in the Quantum Espresso (QE) software package,[2] with the Generalized Gradient Approximation of the exchange-correlation potential and Hubbard U correction (GGA+U)for the strong correlations in 4$f$ bands (6.7 eV for Gd-4$f$) [3] of the exchange-correlation potential with the PBEsol parametrization.[4] Projector-augmented wave (PAW) potentials for Ge, and Ru were taken from the PSlibrary v.1.0.0 set, and for Gd, PAW potential developed by VLab was used.[5] The self-consistencies were carried out using 13 ×13 × 6 $k$-mesh in the irreducible Brillouin zone. Kinetic energy cutoff for charge density and wavefunctions was set to 60 eV and 412 eV. The pseudo potential

DFT calculations for the *J*-coupling constant used the same parameter, except the *k*-mesh was changed to 3 × 3 × 2 to account for the (2*a*,2*b*,*c*) super cell.

**Figure S1.** SEM-EDS results of a $GdRu_2Ge_2$ crystal from arc-melting, showing the agreement between the experimental Gd:Ru:Ge ratio of 1:2:2 and the theoretical ratio of 1:2:2.

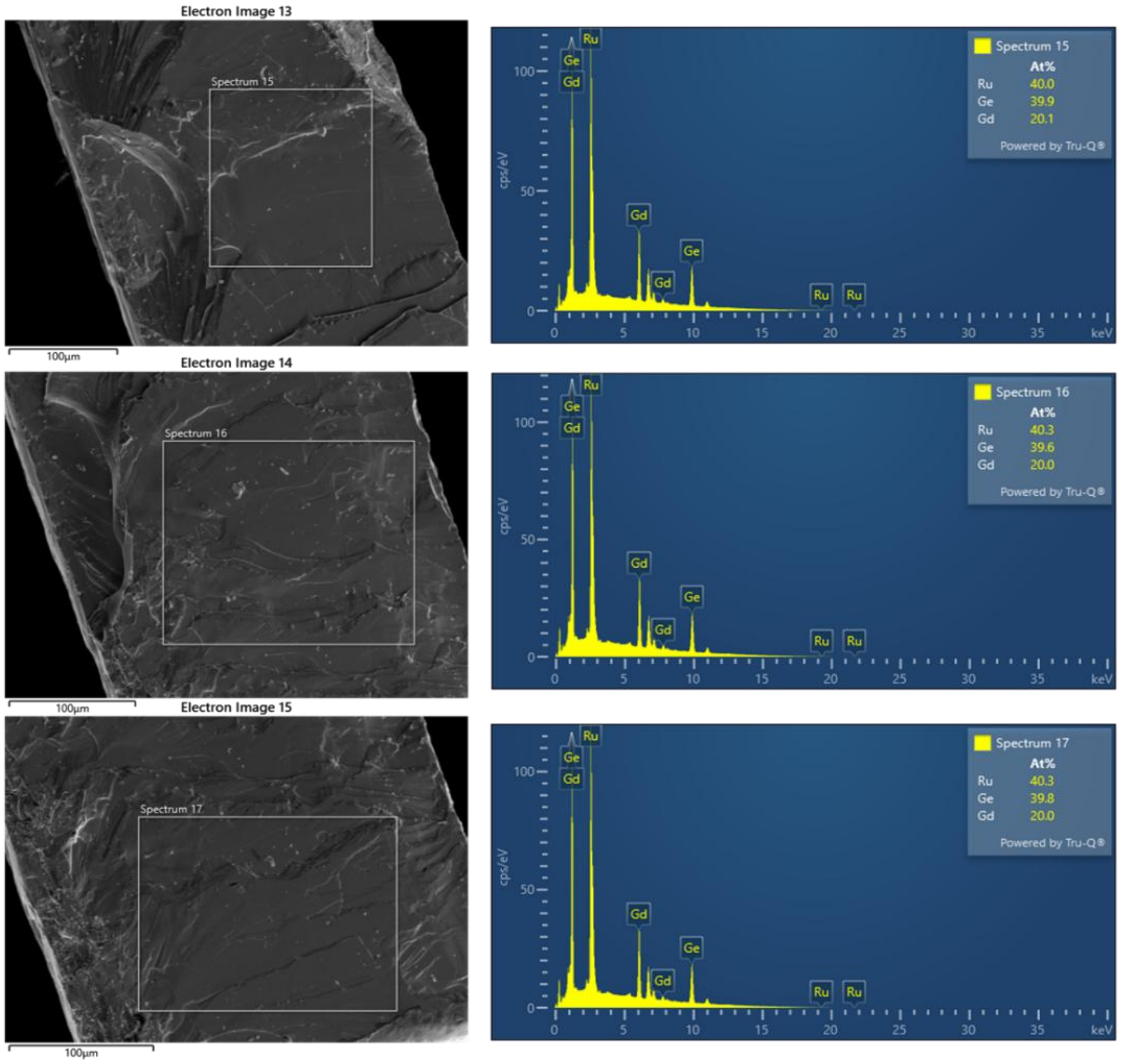

**Figure S2.** (a) $\chi(T)$ at different magnetic fields $\mu_0 H \perp c$, (b) First derivative of magnetization with respect to temperature, $\mathrm{d}M/\mathrm{d}T$ for $\mu_0 H \perp c$, (c) A map of $\mathrm{d}M/\mathrm{d}T$ = $\mathrm{d}S/\mathrm{d}T$ indicating first-order transitions, for $\mu_0 H \perp c$, (d) First derivative of magnetization with respect to the magnetic field $\mathrm{d}M/\mathrm{d}H$ curve $\mu_0 H \perp c$, (e) Isothermal magnetic entropy at each different magnetic field obtained by the integral of $\mathrm{d}M/\mathrm{d}T$ with respect to the magnetic field $\mu_0 H \perp c$.

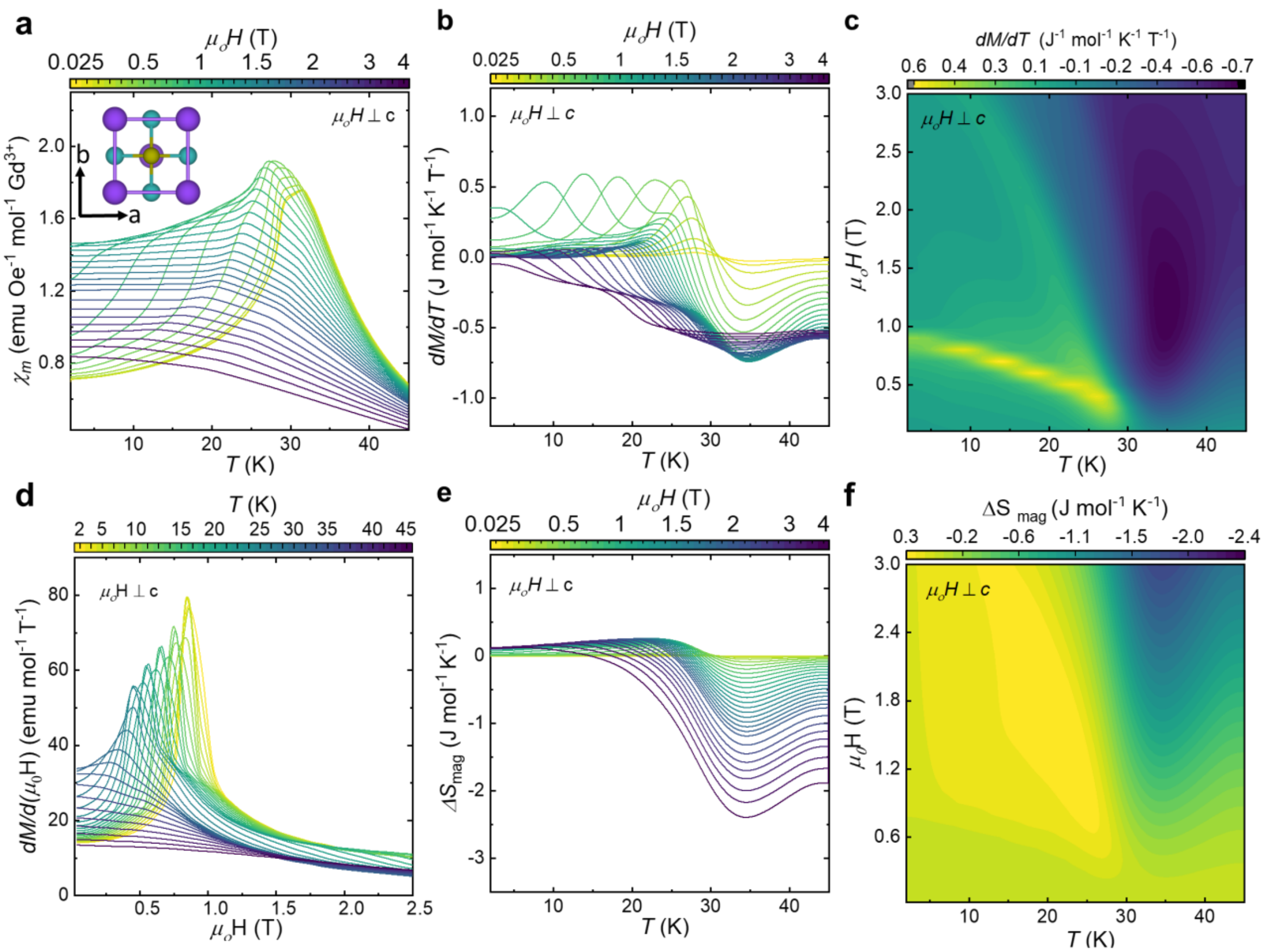

**Figure S3.** Temperature-dependent magnetic susceptibility under constant magnetic field (teal) and Curie-Weiss analysis(black) $\mu_0 H$ // *ab*.

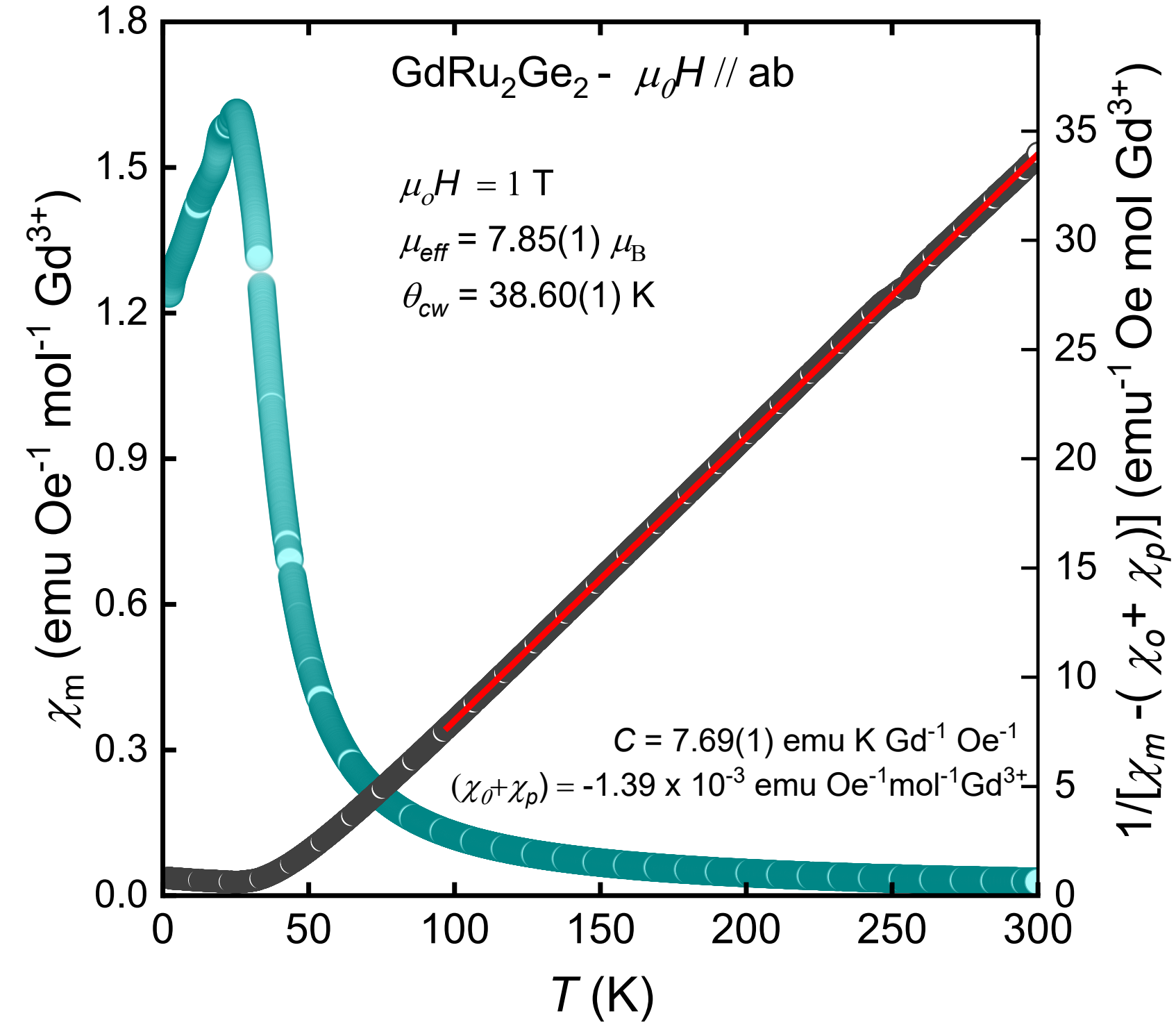

**Figure S4**. (a) Molar heat capacity over temperature ($C_p/T$) vs. Temperature at different fields for powder sample, (b) Molar heat capacity over temperature ($C_p/T$) vs. temperature at $\mu_0 H$ = 0 for $GdRu_2Ge_2$ powder, and magnetic entropy change from 2 K to 300 K, $\Delta S_{mag}$ (2 K ≤ $T$ ≤ 300 K) = 8.5(1) J mol f.u.$^{-1}$ K$^{-1}$ (purple line), expected value for Rln(8) (red line).

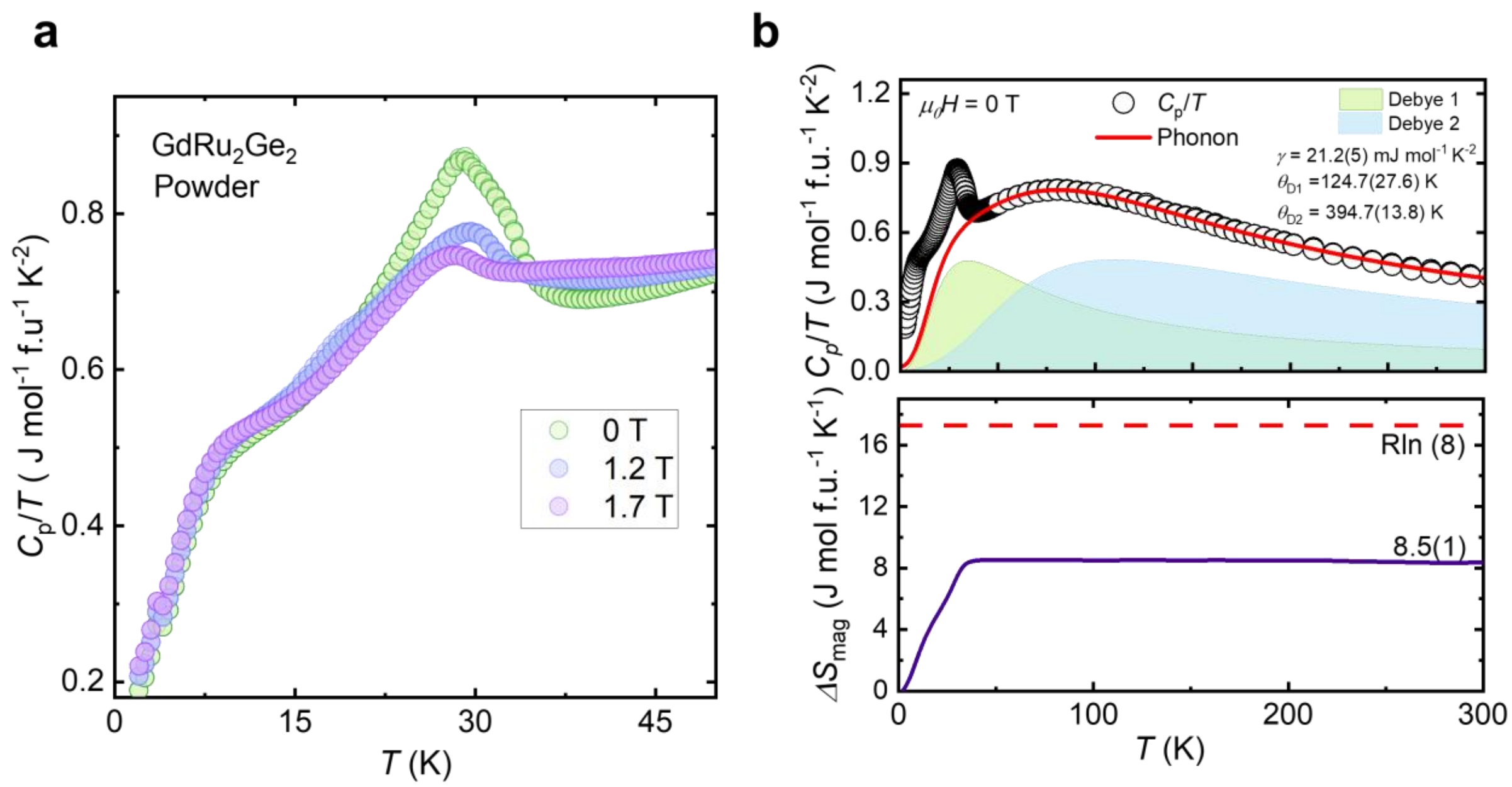

**Figure S5.** $C_p/T^3$ vs log $T$ at zero field indicating the absence of Einstein mode phonons.

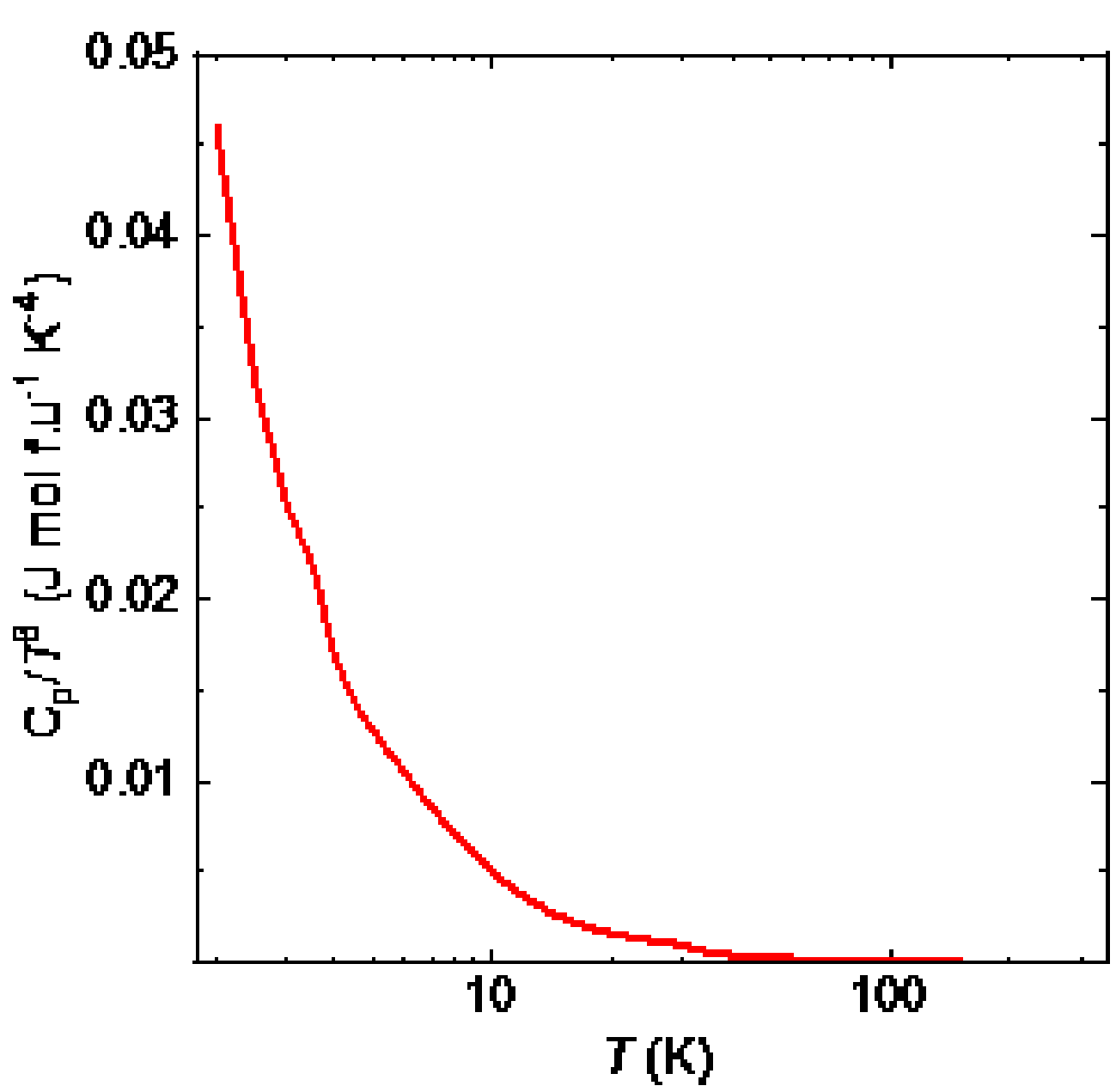

**Figure S6.** Temperature dependence of magnetoresistance in different fields**.**

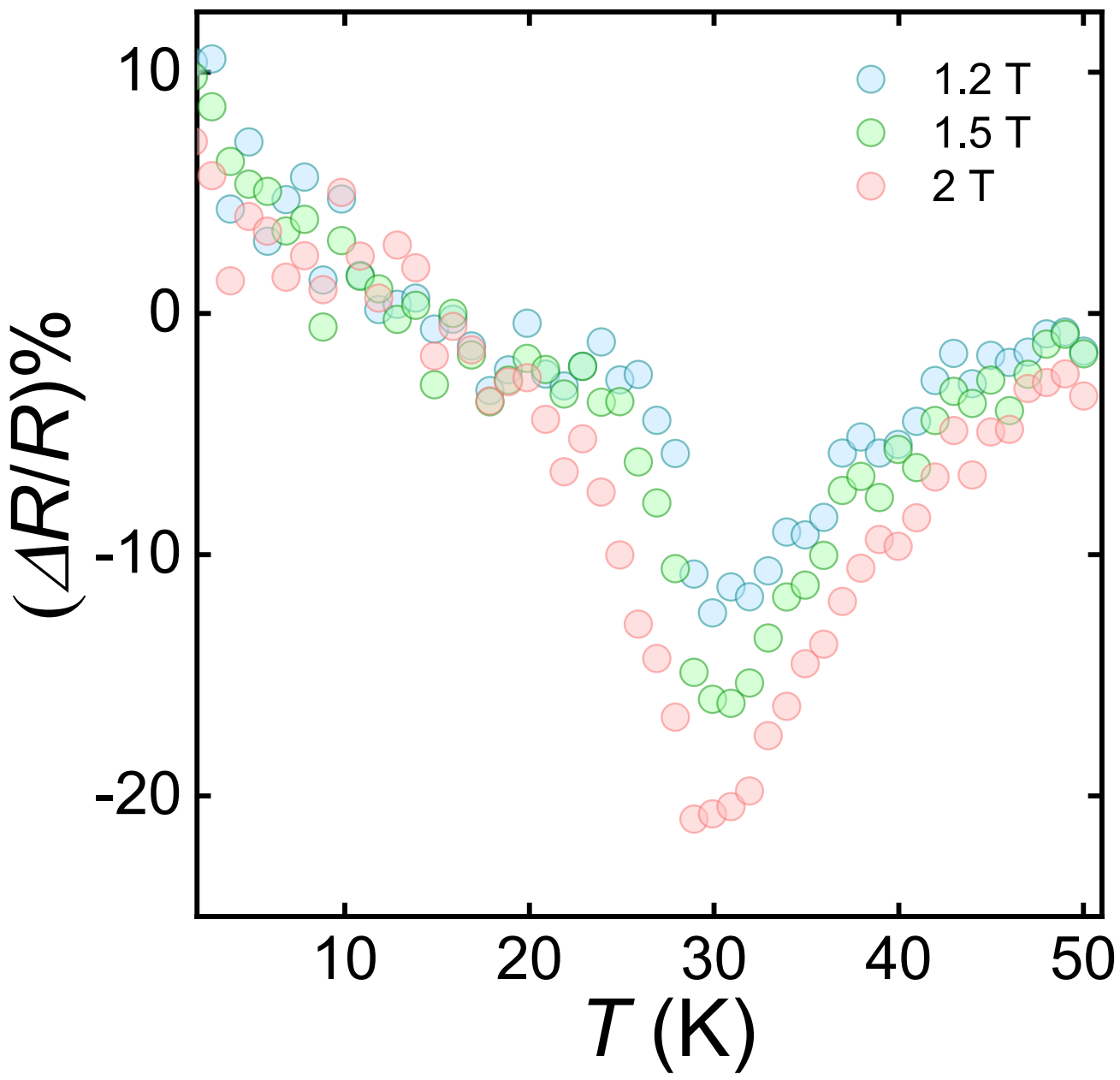

**Figure S7.** Thermogravimetric analysis of $GdRu_2Ge_2$ indicating the congruent melting nature.

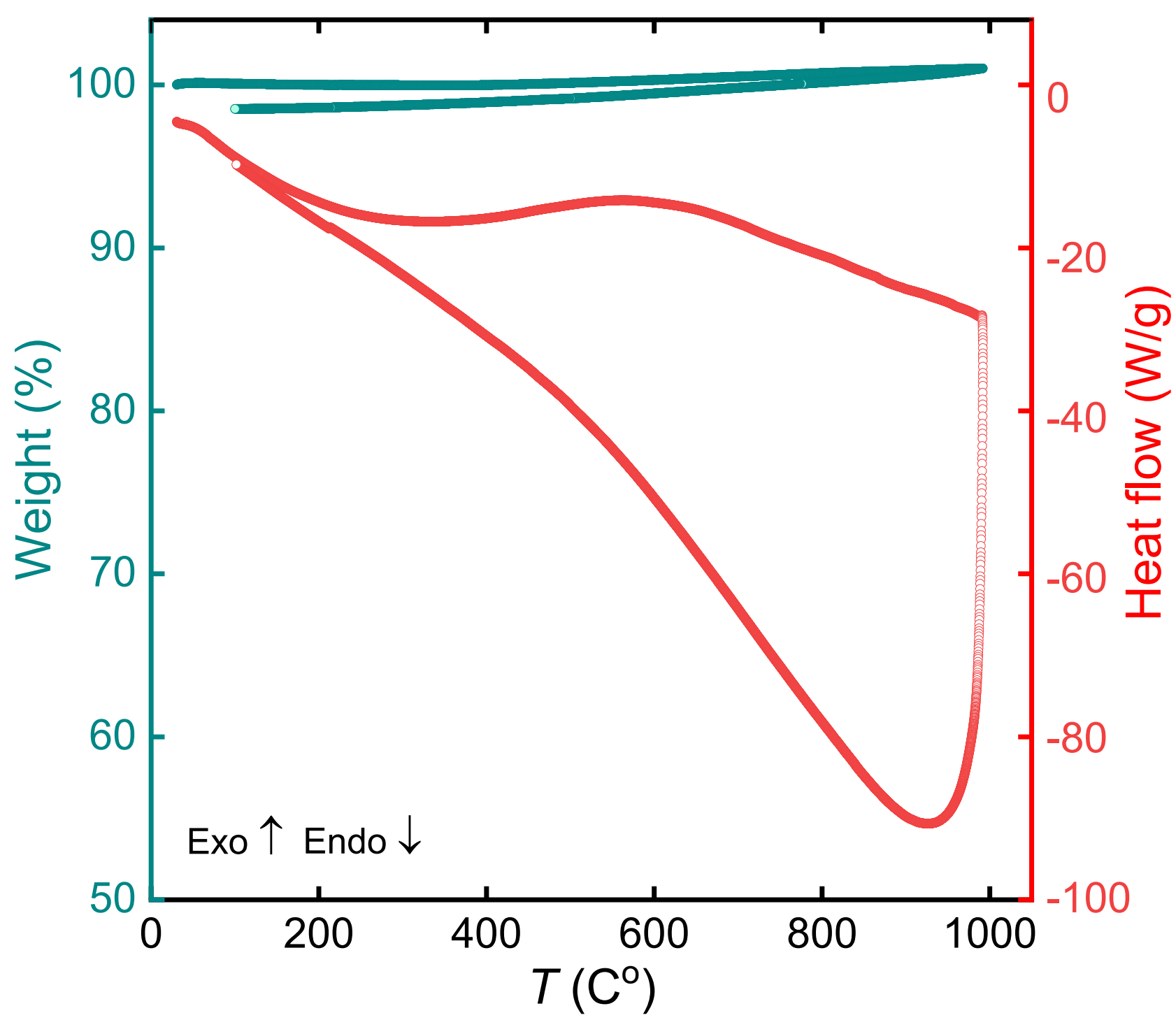

**Figure S8.** Brillouin zone for body-centered tetragonal lattice.[5b, 6]

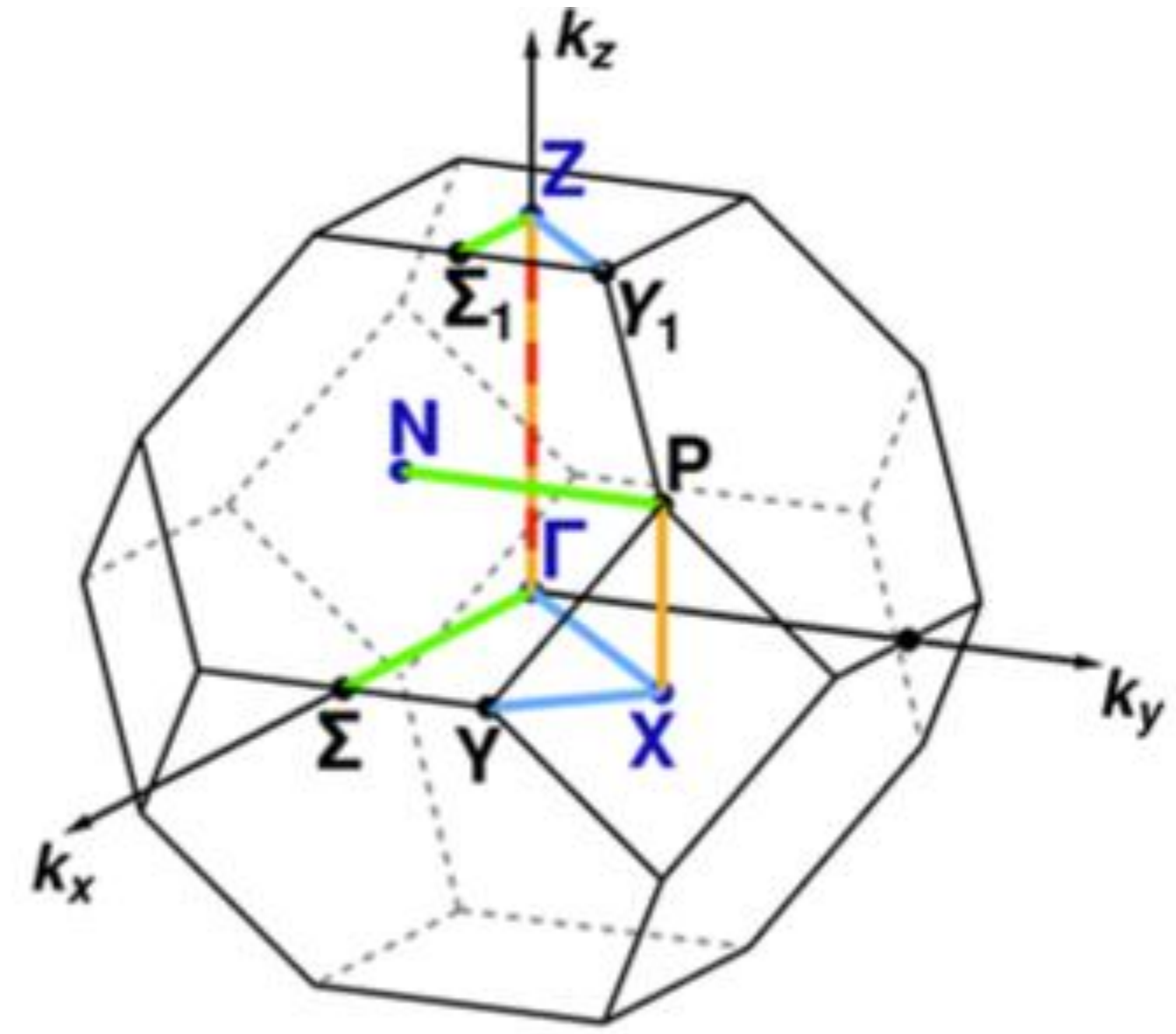

**Figure S9.** Pseudopotential spin-polarized decomposed orbital contributions of $GdRu_2Ge_2$ around the Fermi level.

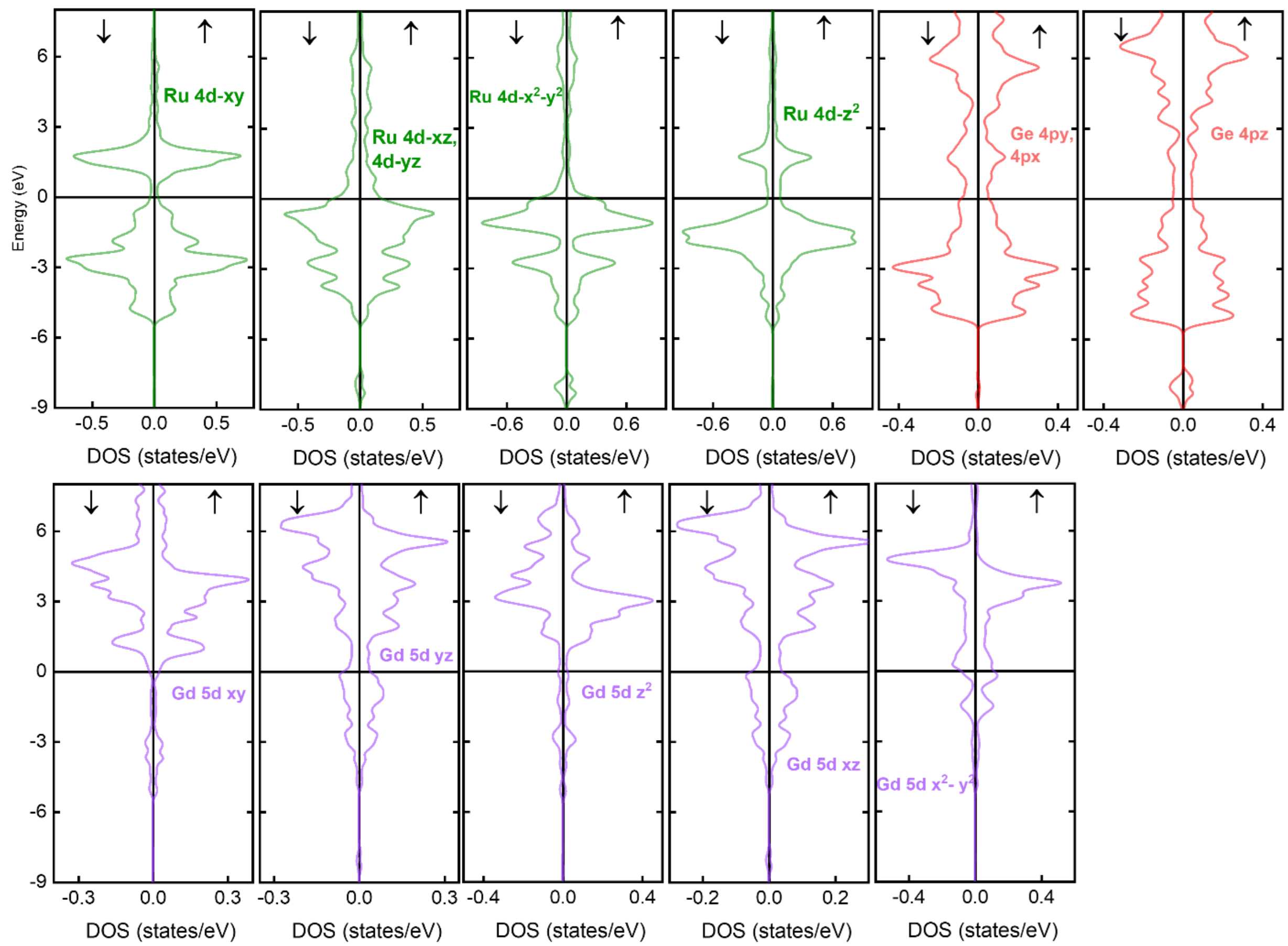

**Figure S10.** Pseudopotential non-spin-polarized density of states of $GdRu_2Ge_2$

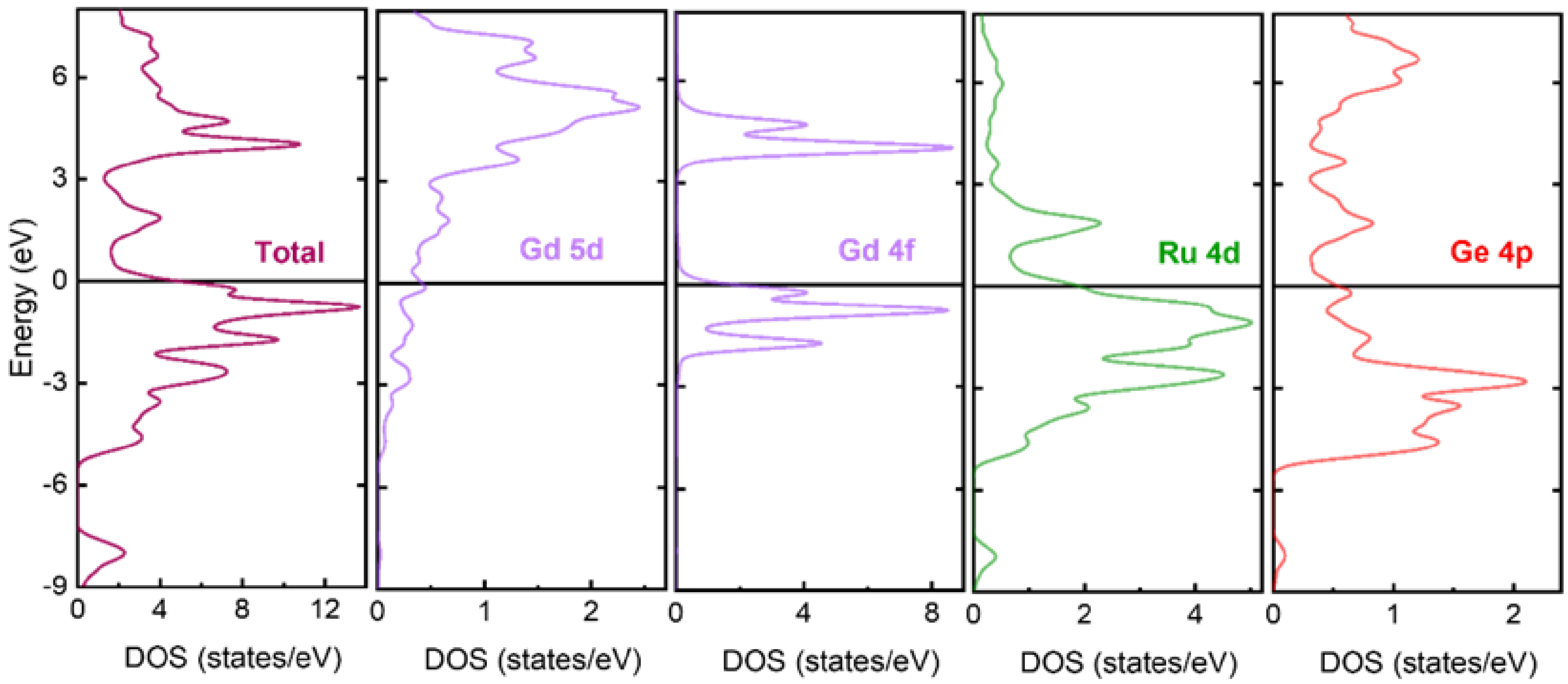

The DFT results without spin-polarization reveal diffuse Gd-4*f* orbital near the Fermi level, which is inconsistent with the localized nature of the 4*f* states, confirming the magnetic properties of the material.

**Figure S11.** DOS showing diffused energy features energy between $GdRu_2Si_2$ and $GdRu_2Ge_2$

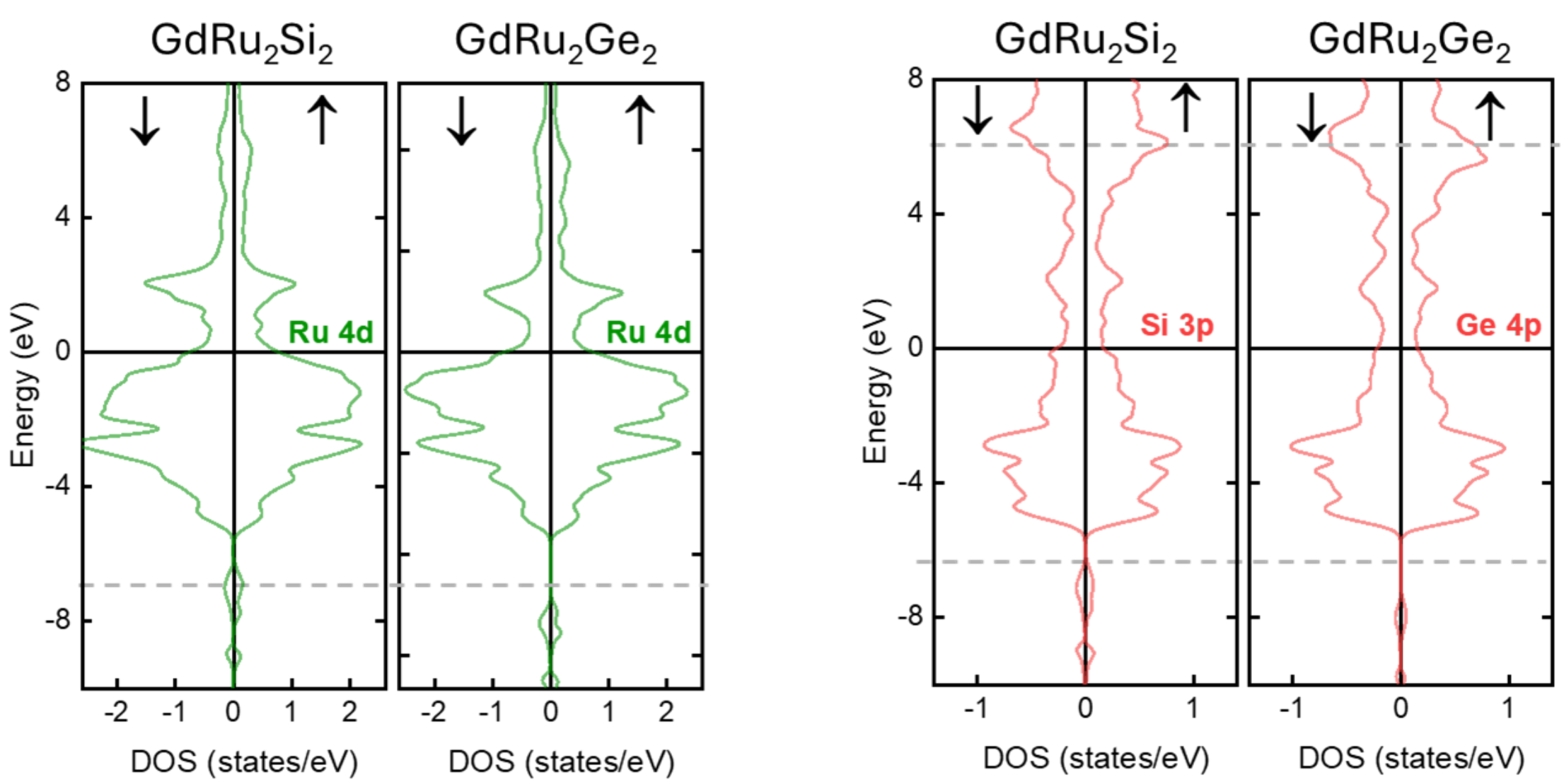

**Table S1.** Crystal structure information and refinement parameters for Gd obtained by single crystal X-ray diffraction.

| | |
|---|---|
| Formula | $GdRu_2Ge_2$ |
| $D_{calc.}$/ g cm$^{-3}$ | 9.486 |
| $m$/mm$^{-1}$ | 43.344 |
| Formula Weight | 504.608 |
| $T$/K | 300(2) |
| Crystal System | tetragonal |
| Space Group | $I4/mmm$ |
| $a$/Å | 4.2316(1) |
| $b$/Å | 4.2316(1) |
| $c$/Å | 9.8797(5) |
| V/Å$^3$ | 176.91(1) |
| $Z$ | 2 |
| Wavelength/Å | 0.71073 |
| Radiation type | Mo $K_a$ |
| $\Theta_{min}$/° | 5.242 |
| $\Theta_{max}$/° | 29.516 |
| GooF | 1.235 |
| $wR_2$ | 0.0323 |
| $R_1$ | 0.0134 |

**Table S2.** Fractional Atomic Coordinates and Equivalent Isotropic Displacement Parameters (Å$^2$) for $GdRu_2Ge_2$. $U_{eq}$ is defined as 1/3 of the trace of the orthogonalised $U_{ij}$.

| **Atom** | **x** | **y** | **z** | $\boldsymbol{U_{eq}}$ |
|---|---|---|---|---|
| Gd1 | 1.500000 | 0.500000 | 0.500000 | 0.0075(2) |
| Ru1 | 1.000000 | 0.500000 | 0.250000 | 0.0053(2) |
| Ge1 | 0.500000 | 0.500000 | 0.13007(8) | 0.0060(2) |

**Table S3.** Bond distances of $GdRu_2Ge_2$

| Bond type | Bond distance (Å) |
|---|---|
| Gd - Gd | 4.2316(1) |
| Ru-Ru | 2.9922(1) |
| Ge-Ge | 2.5701(9) |
| Gd - Ru | 3.2523(1) |
| Gd-Ge | 3.2565(4) |
| Ru-Ge | 2.4250(4) |

**Table S4a.** Phonon estimation using two Debye models (crystal sample)

| Variable | Fitted value |
|---|---|
| Number of oscillators in Debye-1 mode | 3 (Fixed) |
| Number of oscillators in Debye-2 mode | 2 (Fixed) |
| Debye temperature 1 | 457.2 (2.2) K |
| Debye temperature 2 | 175.8 (1.4) K |
| $\gamma$ | 9.6 (1) mJ $mol^{-1}$ $K^{-2}$ |
| $R^2$ | 0.99854 |

**Table S4b.** Phonon estimation using two Debye models (powder sample)

| Variable | Fitted value |
|---|---|
| Number of oscillators in Debye-1 mode | 1.2(2) |
| Number of oscillators in Debye-2 mode | 3.7(2) |
| Debye temperature 1 | 123.4(27.6) K |
| Debye temperature 2 | 394.7(13.8) K |
| $\gamma$ | 21.21(5.1) mJ $mol^{-1}$ $K^{-2}$ |
| $R^2$ | 0.99953 |

**Table S5.** Transverse resistivity fitting parameters and error bar plot

| Temperature (K) | $R_0$ (nOhm·cm/T) | $R_S$ (nOhm·cm$^2$/A) | Carrier Density x $10^{23}$ (1/cm$^3$) |
|---|---|---|---|
| 2 | -2.35 | 0.009 | 2.65 |
| 10 | -3.99 | 0.014 | 1.56 |
| 15 | -3.58 | 0.018 | 1.74 |
| 20 | 0.88 | 0.018 | 7.10 |
| 25 | 1.95 | 0.021 | 3.20 |
| 30 | 5.24 | 0.019 | 1.19 |
| 33 | 5.09 | 0.021 | 1.23 |
| 50 | 0.12 | NA | 0.53 |
| 100 | 8.32 | NA | 0.75 |
| 200 | 7.11 | NA | 0.88 |
| 300 (0-5 T) | -0.47 | NA | 13.2 |
| 300(5-7 T) | -25.7 | NA | 0.24 |

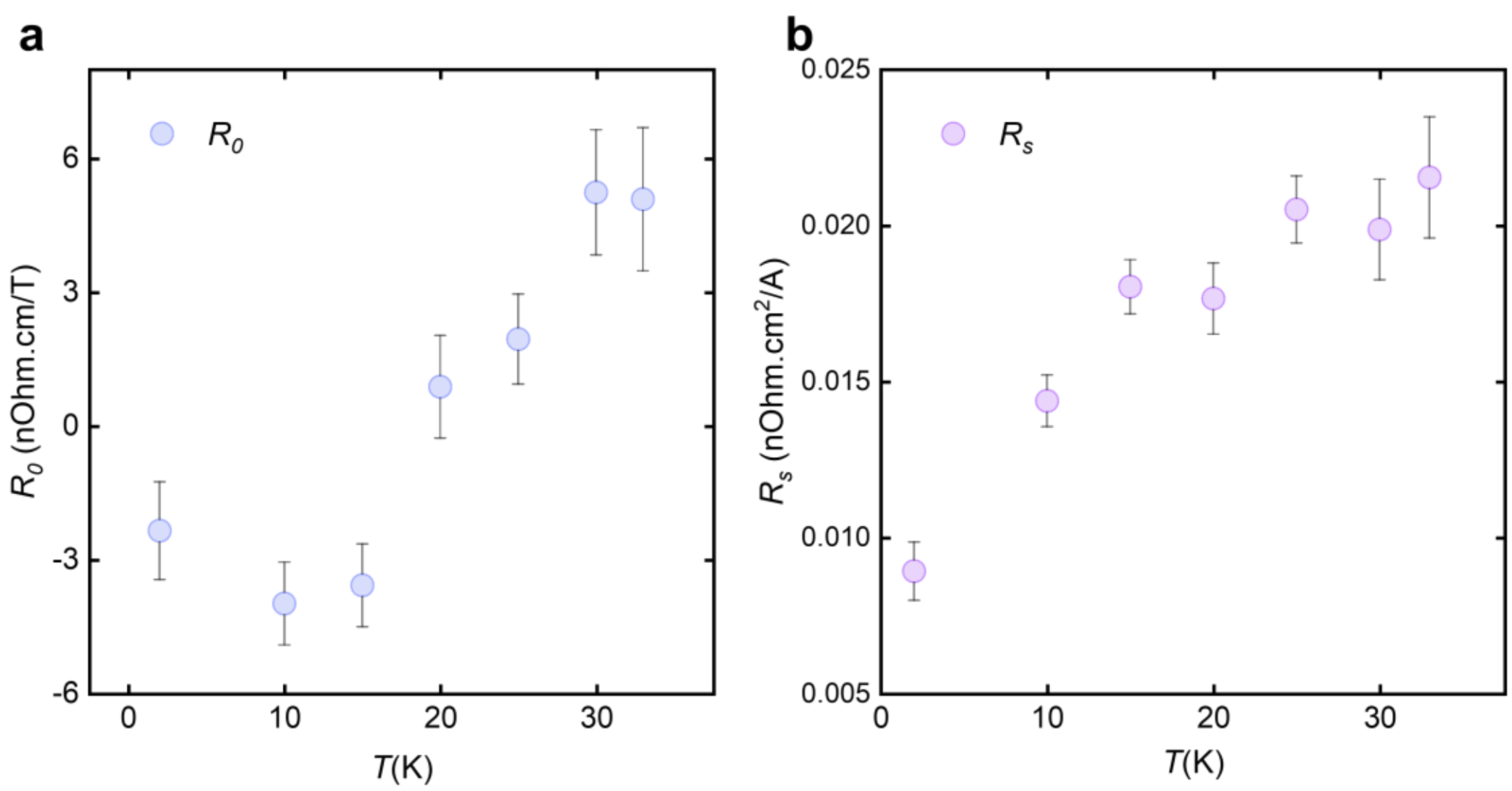

**Table S6.** Integrated crystal orbital Hamilton population (ICOHP) values indicating the bond strength

| **Compound** | **ICHOP value for Gd-Ru bond (eV)** | **ICHOP value for Gd-X bond (eV)** |
|---|---|---|
| $GdRu_2Si_2$ | -1.12266 | -1.14152 |
| $GdRu_2Ge_2$ | -0.99348 | -1.08510 |
| | 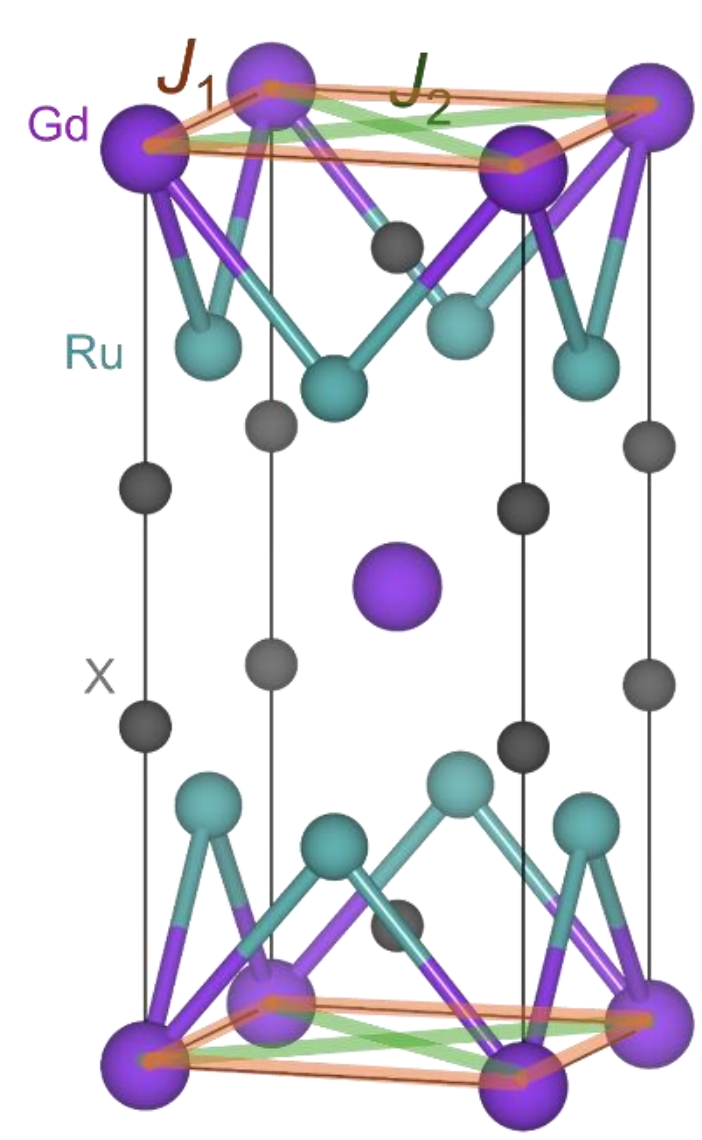  | 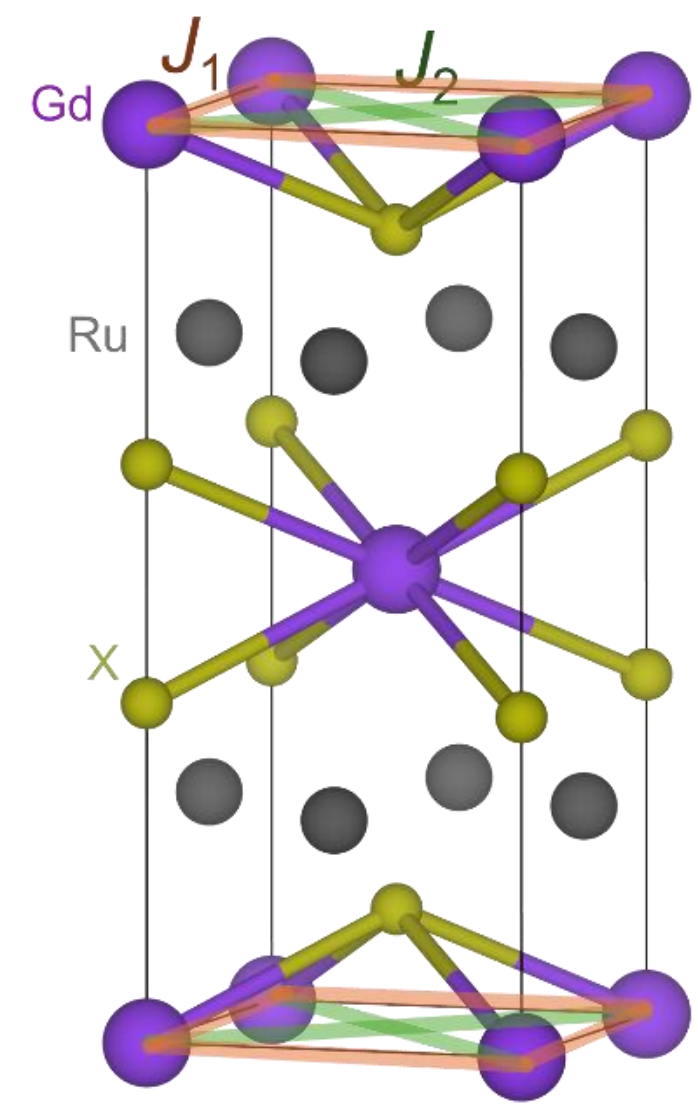  |

**Table S7**. Integrated Crystal orbital overlap (ICOOP) values indicating the orbital overlap.

| Bond type | ICOOP values |
|---|---|
| Ru-Si | 0.0907 |
| Ru-Ge | 0.0980 |
| Gd-Si | 0.0219 |
| Gd-Ge | 0.0247 |